\documentclass[aps,showpacs,pra]{revtex4}
\usepackage{dcolumn}
\usepackage{graphicx}
\usepackage{amssymb}
\usepackage{epstopdf}
\usepackage[english]{babel}
\usepackage[utf8x]{inputenc}
\usepackage{wrapfig}
\usepackage{lipsum}
\usepackage{enumitem}
\usepackage{color,soul}

\begin{document}

\title{document}
\title{Rabi flop-assisted electron tunneling through the quantum dots}
\author{S.~E.~Shafraniuk}
\affiliation{Tegri LLC, 558 Michigan Ave, Evanston, 60202, IL, USA}
\pacs{DOI: 10.1109}

\date{\today}

\begin{abstract}
When the external electromagnetic field (EF) with the frequency $\omega $ acts on the discrete levels in the quantum dot (QD), it induces the mixed quantum state characterized by Rabi flops (RF). The RF process involves the oscillations in the level population $\delta n_{\alpha \beta }\left( t\right) $ accompanied by the cyclical absorption and re-emission of photons. The RF time dynamics depend on the Rabi frequency $\omega _{\mathrm{R}}$ and detuning $\Delta =\omega - \omega_Q $, where $\omega_Q =\left( \varepsilon _{\beta }- \varepsilon _{\beta }\right) /\hbar$, $\varepsilon _{\alpha} $ and  $\varepsilon _{\beta} $ are the level energies. By using the Floquet formalism to solve the time-dependent wave equations for the voltage-biased QD exposed to EF, we examine how RF is expressed in the electric current. We find that the Rabi flop-assisted tunneling (RFAT) is pronounced in the I-V curves as steps and flat plateaus, whose position and spacing reflect the intrinsic features of the mixed quantum state and directly depend on $\omega _{\mathrm{R}}$ and $\Delta $. Thus, measuring RFAT allows an immediate observation of RF right in the I-V curves, thus improving the versatility and accuracy of many applications in the broad frequency range.
\end{abstract}

\maketitle

\section{\label{sec:level1}Introduction}

Further improving the technological potential of the low-dimensional systems to build solid-state THz detectors and emitters of electromagnetic waves is important due to their numerous applications in various areas of science and technology. The performance of the respective devices is characterized by the response time, portability, noise equivalent power (NEP), the dependence of it on the temperature, etc \cite{Dereniak,HaugKoch,Levi-book,Cohen-T}. The optimization of the above parameters requires a deep knowledge of the interaction mechanisms between such devices and the external electromagnetic field. Such interaction frequently involves fundamental quantum effects creating completely new ways of electronic transport in low-dimensional systems such as quantum dots (QD) \cite%
{Kawano1,Rinzan,Kawano2,Mele,Vitiello1,Vitiello2,My-JP11,My-EPJ,My-JP8,My-PRB8,My-PRB7,My-Nanotech}. 
In this respect, an increased interest attracts the simplest case when the effective Hilbert space of QD is reduced to just a few quantum states. Understanding the interplay between electron transport and the driving field in QD is of utmost importance, both from the fundamental and applied points of view. In particular, the ability to rapidly measure and control electron states using ac fields has immediate applications to quantum metrology and quantum information processing. Therefore, the electron transport through QD exposed to the electromagnetic field (EF) is a subject of intensive experimental and theoretical research \cite%
{Kawano1,Rinzan,Kawano2,Mele,Vitiello1,Vitiello2,My-JP11,My-EPJ,My-JP8,My-PRB8,My-PRB7,My-Nanotech} The study\cite%
{Kawano1,Rinzan,Kawano2,Mele,Vitiello1,Vitiello2,My-JP11,My-EPJ,My-JP8,My-PRB8,My-PRB7,My-Nanotech} had been motivated by an interest in the nanoscale phenomena which are exploited in various sensors and detectors of external fields and substances \cite{Dereniak,HaugKoch,Levi-book,Cohen-T}. 

\begin{figure}
\includegraphics[width=85 mm]{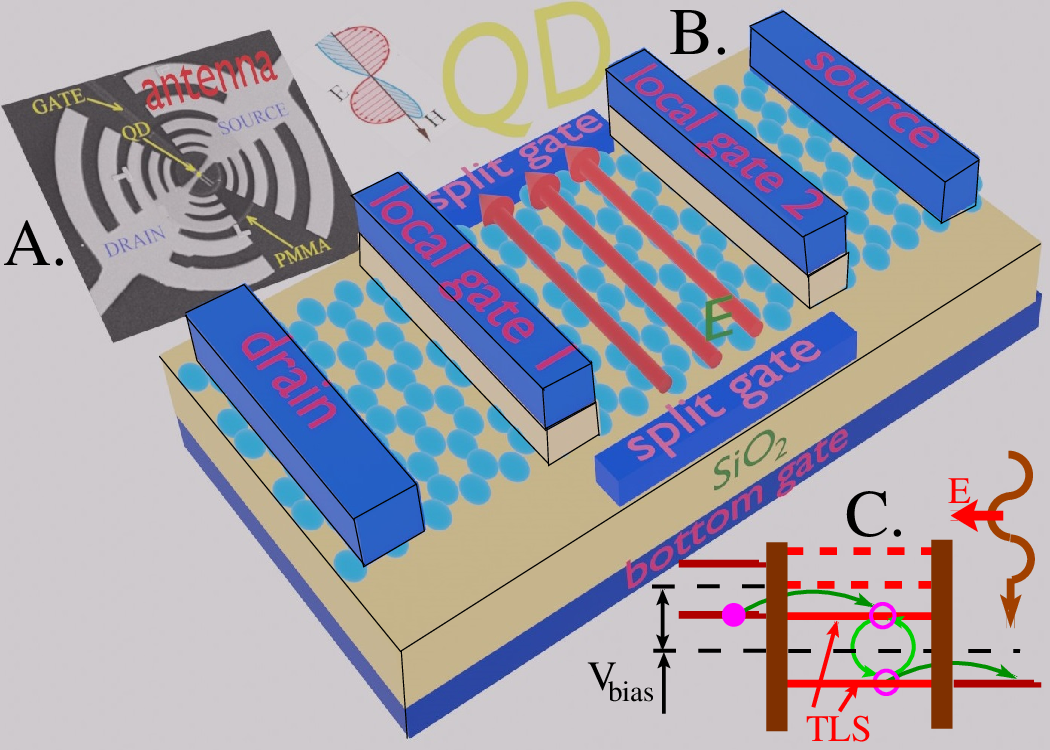} 
\caption{{The electrically controlled QD is formed on a graphene stripe with zigzag atomic edges (ZZ-stripe). The transversal electric field ${\bf E}$ (red arrows) applied between the split gates induces the dc Stark splitting $\Delta_{\rm dc} $ of the energy levels to create TLS. The bottom gate controls the mean value of the electrochemical potential $\mu $ in the ZZ-stripe while the local gates tune the height of the interdot chiral barrier $V_{\rm B}$. A. The bowtie antenna where the QD location is in the center. B. The FET device where the voltage-biased QD is formed on the graphene stripe with zigzag edges. The bottom gate sets the average electrochemical potential $\mu $, the two local gates control the QD barrier strength, and the split gate controls the TLS level spacing. C. The energy diagram of QD where the Rabi flops in TLS are probed using the Rabi flop-assisted tunneling (RFAT). Red dash lines show the pseudoenergies $W_k$ defined by Eqs.~(s4), (s6) in Supplementary material.}}
\label{Fig_1}
\end{figure}

To probe or generate EF, many existing infrared and THz devices, such as qubits, detectors, and lasers measure the electric current flowing through the quantum dots  \cite{Kawano1,Rinzan,Kawano2,My-Nanotech}. In Fig.~\ref{Fig_1} we show QD based on the field effect transistor (FET) that is formed by an isolated central island where the attached source, drain, and gate electrodes control the quantized electron states by applying respective electric potentials. By utilizing the all-electric control of FET one accomplishes fine-tuning the interaction between EF and QD \cite{My-PRB19,My-AQT,My-PRB23}. Similar QD devices are used in experimental physics, nanoelectronics, chemistry, industry, and medicine \cite{Asada,Kitagawa,Arzi,Wilmart2020,Shaikhai,Island,Dyak}.  

In quantum dots, the interaction between EF and QD results in the photon-assisted tunneling (PAT) \cite{Dayem-Martin,Tien-Gordon,Platero,Brune}. It also causes Rabi flops \cite{Bloch,Viebahn,Broers,Merlin} when the oscillations of the population of energy levels are accompanied by cyclic absorption of the photons, and stimulate re-emitting them back. The period and shape of such oscillations depend on the Rabi frequency $\omega _{\mathrm{R}}$ and detuning $\Delta =\omega -\omega _{Q}$, where $\omega $ is the EF angular frequency, $\omega _{Q} = \left(\varepsilon _{\alpha }-\varepsilon _{\beta }\right) /\hbar $, $\varepsilon_{\alpha}$ and $\varepsilon_{\beta}$ are the quantized level energies with indices $\alpha $ and $\beta $. The Rabi frequency $\omega _{\mathrm{R}}$ is a semi-classical concept and is associated with the strength of the coupling between light and the transition that determines the time scale at which the interlevel transition proceeds. By definition, , $\omega _{\mathrm{R}}=\left( \mathbf{d}\cdot \mathbf{E}_{\mathrm{ac}}\right) /\hbar $ is the Rabi frequency, $\mathbf{d}$ is the transition dipole moment, $\vert \mathbf{E}_{\mathrm{ac}}\vert = V_{\mathrm{ac}}/W $ is the amplitude of the electric field and $W $ is the stripe width.  For the two-level system (TLS) there is a simple analitical expression (see, e.g., Ref.~\cite{Hirschfelder1}) $\omega_{\rm R} = \sqrt{\Delta^2 + \varsigma^2}$, where $\varsigma $ is the inter-level coupling strength. Rabi flopping occurs with frequency $\omega _{\mathrm{R}}$ between the energy levels illuminated with resonant light. The occurrence of Rabi flops in solid-state QD typically is suppressed when the level of noise is high while the coherence and relaxation rates are big. That complication can be solved by exploiting the spectral narrowing of quantized levels by several orders of magnitude \cite{Asada,Kitagawa,Arzi,My-PRB23} by using a device similar as sketched on Fig.~ \ref{Fig_1} in Sec.~\ref{sec:level3}.

In this paper, we study electronic transport through quantum dots where the peculiar synergism between ac fields and quantum confinement gives rise to novel phenomena. On the one hand, in a typical PAT process, the electrons tunneling through the potential barrier absorb photons of external field with frequency $\omega $ resulting in ascending steps on the IV curve spaced by $\delta I_{\rm PAT} = \hbar \omega/(e R_{\rm T})$. On the other hand, in Rabi flops, the oscillations of the population of energy levels result in the emission/absorption of photons with a Rabi frequency $\omega _{\mathrm{R}}$. The latter PAT process here is the Rabi flop-assisted tunneling (RFAT), whose microscopic mechanism involves the mixed quantum state as explained below. As we shall in Sec.~\ref{sec3-3}, the Rabi flops cause respective ascending/descending steps on the IV curve, whose spacing is $\delta I_{\rm RFAT} = \hbar \omega _{\mathrm{R}}/(e R_{\rm T})$. We emphasize that $\delta I_{\rm RFAT} \neq  \delta I_{\rm PAT} $ because the involved physics  differs.

The model is outlined in Sec.~\ref{sec2}, where in Sec.~\ref{sec:level2} we discuss known mechanisms of PAT  \cite{Dayem-Martin,Tien-Gordon,Platero}. In Sec.~\ref{sec2}~B we introduce the prototype device represented in Fig.~\ref{Fig_2}. In Sec.~\ref{sec2-2} we outline the approach and show the process flow diagram in Fig.~\ref{Fig_3}. To describe RFAT in QD, we use the Floquet formalism to solve the time-dependent electron wave function equations [see Sec.~\ref{Supplement}] and compute the electron spectrum (Sec. A), the density matrix  (Sec. B), and the electric current  (Secs. C, D) in Sec.~\ref{Supplement}. In Sec.~\ref{sec3} we present the obtained results illustrating RFAT for the simplest two-level system (TLS) exposed to EF and described by the Floquet method.  Using the Floquet formalism to solve the time-dependent wavefunction equations, we compute the electric current in QD. In Sec.~\ref{sec3-1} we compute the time-dependence of the TLS density matrix $\rho _{\alpha \beta } (t)$ for various EF frequencies, dipole coupling strength, and detuning. In Sec.~\ref{sec4} we establish the connection between $\rho _{\alpha \beta } (t)$ and the time-dependent oscillatory change $\delta n_{i}(t) $ in the distribution function $n_{i} = n_{i}^{\left( 0\right) }  +  \delta n_{i}(t) $ related to the single-particle energy eigenstate $\varepsilon _{i}$ \ localized in QD. This relates the Rabi flops to the non-stationary changes of the electron distribution function $\delta n_{i}(t) $. In Sec.~\ref{sec3-3} we present the calculation results based on our model of RFAT outlined in the former Sec.~\ref{sec2-2}. We see that the RFAT is pronounced in the I-V curves as a series of steps and flat plateaus spaced by $\Delta  V_{\mathrm{bias}}^{\mathrm{RFAT}}=\hbar \omega _{\mathrm{R}}/e $, which differ from the conventional Dayem-Martin PAT steps~\cite{Dayem-Martin,Tien-Gordon,Platero} spaced by $\Delta V_{\mathrm{bias}}^{\mathrm{PAT}}=\hbar \omega /e$, since $\omega \neq \omega _{\mathrm{R}}$. In Sec.~\ref{sec3-3} we consider the case when EF is tuned in resonance (or close) to the transition frequency of a given spectral line associated with quantized states in QD. This creates conditions for the Autler--Townes effect \cite{AutTow} (also known as the AC Stark effect or a dynamical Stark effect), causing changes in the shape of the absorption/emission spectra. In Sec.~\ref{sec3-7} we find that the ac field acting on QD shifts the strongly coupled bare QD energy eigenstates into two states $\left\vert +\right\rangle $ and $\left\vert -\right\rangle $ separated by $\hbar \omega_{R} $. Therefore,  due to the Autler-Townes splitting, the QD's I-V curve shows two shoulders that are separated by $\hbar \omega _{R}/e$ around the steady-state position. In Sec.~\ref{sec3-6} we present results for the power spectral density and for TLS with the  Autler-Townes splitting. In Sec.~\ref{sec5a} we discuss the obtained results and how to decode the structure of the I-V curve indicating intrinsic features associated with Rabi flops in the course of the EF-induced mixing of the different quantized states in QD. We also outline how to describe the multiple quantum transitions for the two-state system. This allows extracting more of the EF parameters immediately from the I-V curves thus improving the versatility and accuracy of the quantum dot THz   devices used in various infrared and THz applications \cite%
{Kawano1,Rinzan,Kawano2,Mele,Vitiello1,Vitiello2,My-JP11,My-EPJ,My-JP8,My-PRB8,My-PRB7,My-Nanotech,My-PRB19,My-AQT,My-PRB23}. As an experimental prototype, we discuss the all-electrically controlled QD presented in Figs.~\ref{Fig_1}, \ref{Fig_2}, as described below. We conclude that the I-V curve measurement opens the path to an immediate experimental observation of intrinsic features of the mixed quantum state in QD as well as to improve the analysis of the EF spectra in the far infrared and THz frequency ranges.

\section{The model}\label{sec2}

\subsection{\label{sec:level2}Rabi flop-assisted electron tunneling}

The authors \cite{Kawano1,Rinzan,Kawano2} experimentally measured the differential conductance of the CNT quantum dot exposed to the external THz irradiation. They observed considerable changes in the junction$^{\prime }$s conductance due to the photon-assisted tunneling (PAT) of single electrons \cite{Kawano1,Rinzan,Kawano2} regarded as the Dayem-Martin effect \cite{Dayem-Martin,Tien-Gordon,Platero,Brune}. The previous works \cite{Dayem-Martin,Tien-Gordon,Platero,Brune} focused mostly on PAT involving single-electron quantum states. Then the PAT IV curve has the form of an ascending ladder illustrating absorption of the external field photons with frequency $\omega $. For QD, the authors of Ref.~\cite{Brune} obtained the Bessel-type sidebands at separations $n \omega$, where $\omega $ is the external ac field frequency, and $n = \pm 1, \pm 2, \pm 3..., $, etc. However, such subbands are not related to Rabi flops. The correct answer~\cite{Bloch,Viebahn,Broers,Merlin} is well known: the non-stationary effect and the sidebands’ separation are determined by the Rabi frequency $\omega_{\rm R} $, but not by the external ac field frequency $\omega \ne \omega_{\rm R} $, as was stated in Ref.~\cite{Brune}. Furthermore, the rotating-wave approximation (RWA) used in Ref.~\cite{Brune} incorrectly describes the mixed quantum state originating from the polarization by the external ac field of the system with quantized levels. In particular, RWA omits important processes, e.g., the emission of photons with the Rabi frequency $\omega_{\rm R} $. Another disadvantage is using the tunneling Hamiltonian approach, whose validity is questionable but for the low-transparent barriers. 

To describe RFAT we exploit the exact Floquet solution for TLS as described in Secs.~A, B of Sec.~\ref{Supplement}. We use the Floquet solution to compute the electric current through QD, where we single out RFAT, which is a part of PAT but related to the mixed quantum state in TLS. In this way, we emphasize that RFAT is a different component of PAT arising due to the special microscopic mechanism involving the Rabi flop processes. On the one hand, the conventional PAT is typically related to the photon-assisted microscopic process when the tunneling occurs between the left/right single-electron states. On the other hand, RFAT is a sort of PAT meaning that RFAT is caused by Rabi flops attributed to the mixed quantum state when the external field polarizes the quantized states. In either case, there is PAT observable as a distinctive ladder in the IV curves. The differences between the conventional PAT and RFAT are: (a) typically, the photon-assisted tunneling in conventional PAT occurs between the single-electron states, while in RFAT it involves the mixed quantum state in the quantum dot. (b) The photon frequency in PAT coincides with the external field frequency omega, while the photon frequency in RFAT is the Rabi frequency. They are different quantities; (c) In the conventional PAT ladder, the IV-curve steps are ascending (i.e., the external field photons are absorbed during the electron tunneling process) while in the RFAT ladder, depending on the strength and frequency of the ac field, the Rabi photons can either absorbed or emitted resulting in a mixture of ascending or descending steps.

When the electromagnetic field (EF) is applied to QD, there are the following known mechanisms \cite{My-PRB19,My-AQT,My-PRB23} affecting electron transport. (i) Due to the absorption of the EF photons, the electron distribution function inside QD oscillates with time resulting in an additional change in the QD transparency.  (ii) EF causes the high-frequency modulation of the QD's barrier height, thereby modifying the QD electron spectrum and inducing harmonics of the tunneling electric current. It alters the overall electron transmission probability through QD. (iii) The ac field induces the photon-assisted tunneling \cite{Dayem-Martin,Tien-Gordon,Platero} expressed as steps on the I-V curve, whose shape and position indicate the EF parameters. The above physical mechanisms (i)-(iii) are currently exploited for designing the QD-based sensors and lasers acting in a broad spectrum of EF.

\subsection{\label{sec:level3}The prototype device}

\begin{figure}
\includegraphics[width=85 mm]{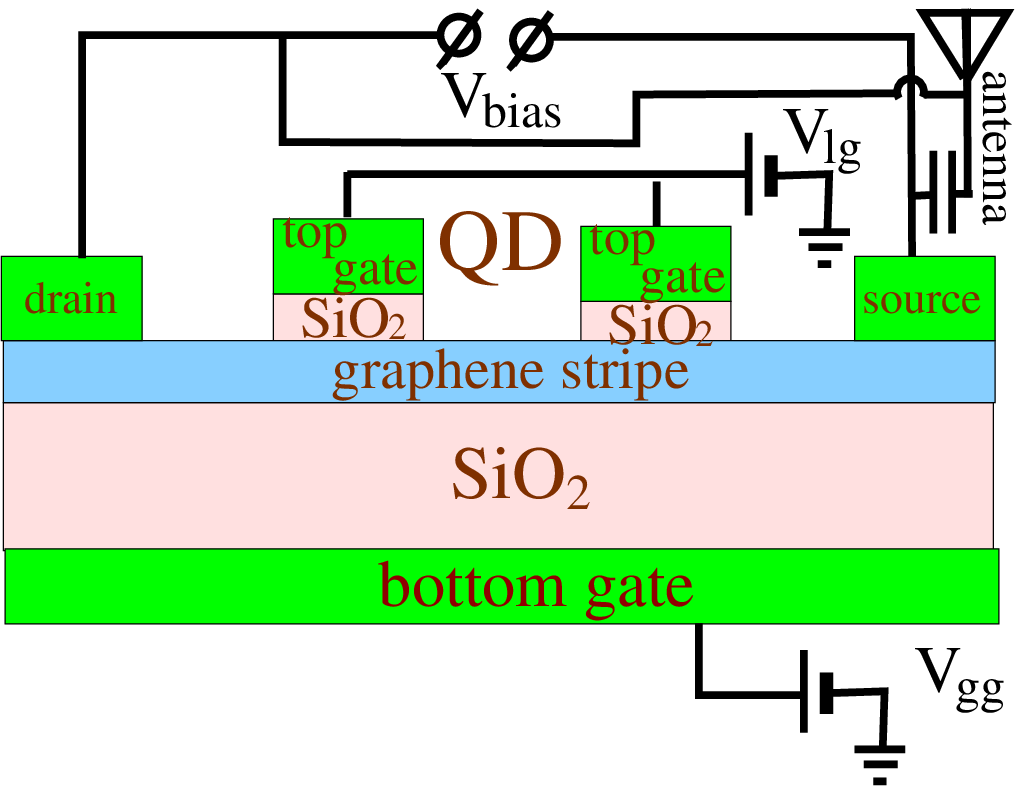} 
\caption{ {The schematic of electric control of the QD cluster, which is depicted in the previous Fig.~\ref{Fig_1}. The I-V curve of electric current between the source and drain electrodes serves to probe the QD spectrum and the Rabi flop dynamics. The bias voltage $V_{\rm bias}$ is applied between the source and drain electrodes to induce the dc electric current in QD. In addition, the ac electric current is induced in QD by the external EF detected by the bowtie antenna, which is capacitively coupled to the source electrode. For simplicity, we do not show here the split-gate voltage $V_{\rm sg} = \vert \mathbf{E}_{\mathrm{sg}}\vert W $, which causes the dc Stark splitting $2\Delta_{\rm dc} = eV_{\rm sg} $ of the energy levels. The $V_{\rm gg} $ voltage is applied to the bottom gate and actually controls the mean value of the electrochemical potential $\mu $ in the ZZ-stripe. The $V_{\rm lg}$ voltage is applied to the local gates to tune the height of the interdot barriers.}}
\label{Fig_2}
\end{figure}

The RFAT mechanism is general and acts in arbitrary voltage-biased QD with quantized electron states exposed to an EF. However, in QD based on conventional semiconductors, the quantum coherence is typically suppressed due to the decoherence, dephasing, and big relaxation rates. The negative impact of the mentioned factors can be reduced by several orders of magnitude using spectral narrowing as suggested in Refs. \cite{Asada,Kitagawa,Arzi}. Another possible prototype device is the quantum dot cluster formed on the narrow graphene stripe, as sketched in Figs.~\ref{Fig_1}, \ref{Fig_2}. According to Ref.~\cite{My-PRB23}, in the graphene QD cluster consisting of one or several QD one achieves the intrinsic spectral narrowing (ISN) of quantized levels by 4-6 orders of magnitude. In either case, spectral narrowing can reduce the negative impact of decoherence, dephasing, and energy relaxation by many orders of magnitude. This results in a great improvement in the functionality of QD devices. In the size of graphene QD cluster is just a few nm, the electrons undergo spatial quantization that is all-electrically tunable  \cite{My-PRB19,My-AQT,My-PRB23} as shown in Fig.~\ref{Fig_1}. Remarkably, in the electron spectrum of the stripe with zigzag atomic edges (ZZ-stripe), there is a very sharp edge level at energy $\varepsilon_{\alpha} =0$. When the transverse electric field $\mathbf{E}_{\mathrm{sg}}$ is applied using the split gates as shown by red arrows in Fig.~\ref{Fig_1}, the single edge level undergoes the dc Stark splitting $2 \Delta_{\rm dc} = e\vert \mathbf{E}_{\mathrm{sg}}\vert W$ into two very narrow and sharp F-levels \cite{Shafr-Graph-Book,TEbook,My-AQT,My-PRB23}. The other quantized P-levels are much weaker and in our model are disregarded. Then, the two arising F-levels can effectively be regarded as the two-level system (TLS), whose properties are controlled by the gate voltages $V_{\rm sg}$ and $V_{\rm gg}$ \cite{My-JP8,My-PRB19,My-AQT,My-PRB23,Shafr-Graph-Book,TEbook}. Furthermore, in such an electrically controlled QD, the electron-phonon coupling is diminished, and the temperature effects are minimized \cite{My-PRB23}. 

The electron transport through QD is initialized by external fields. Relevant microscopic processes involve the penetration of particles from attached electrodes into and out of the dot, inter-electrode transmission of particles through the dot via discrete energy levels, and the field-induced inter-level transitions. When a gate voltage $V_{\mathrm{gg}}$ is applied bottom gate electrode as shown in Fig.~\ref{Fig_2} it functions as a field effect transistor. The electric current is initialized by a source-drain bias voltage $V_{\mathrm{bias}} $ imposed between the source and drain electrodes. In addition, QD is subjected to an electromagnetic field (EF) delivered by antenna that is capacitively coupled to the source and drain electrodes as shown in Fig.~\ref{Fig_2}. Alternatively, antenna may also be coupled to the local gate electrodes or to the split gate electrodes. The Schottky barriers I originate from a difference in concentrations of the charge carriers between the source and drain electrodes and the dot. The gate voltage $V_{\mathrm{gg}} $ sets the value of electrochemical potential $\mu $ of  electrons in QD.

Basically, the reported RFAT mechanism works for arbitrary voltage-biased QD with quantized electron states exposed to EF. In this paper, for definiteness, we assume that the relevant setup represents QD, which is based on the graphene stripe with zigzag atomic edges. Initially, we consider QD hosting the two well-defined electron levels originating from so-called perfectly flat F-bands controlled by the split-gate voltage  \cite{My-AQT,My-PRB23}. The other levels, corresponding to P-bands are much weaker and they can be neglected. Then we consider the more complex situation when Autler-Townes splitting increases the number of levels from two to four. 

\subsection{The Approach}\label{sec2-2}

\begin{figure}
\includegraphics[width=85 mm]{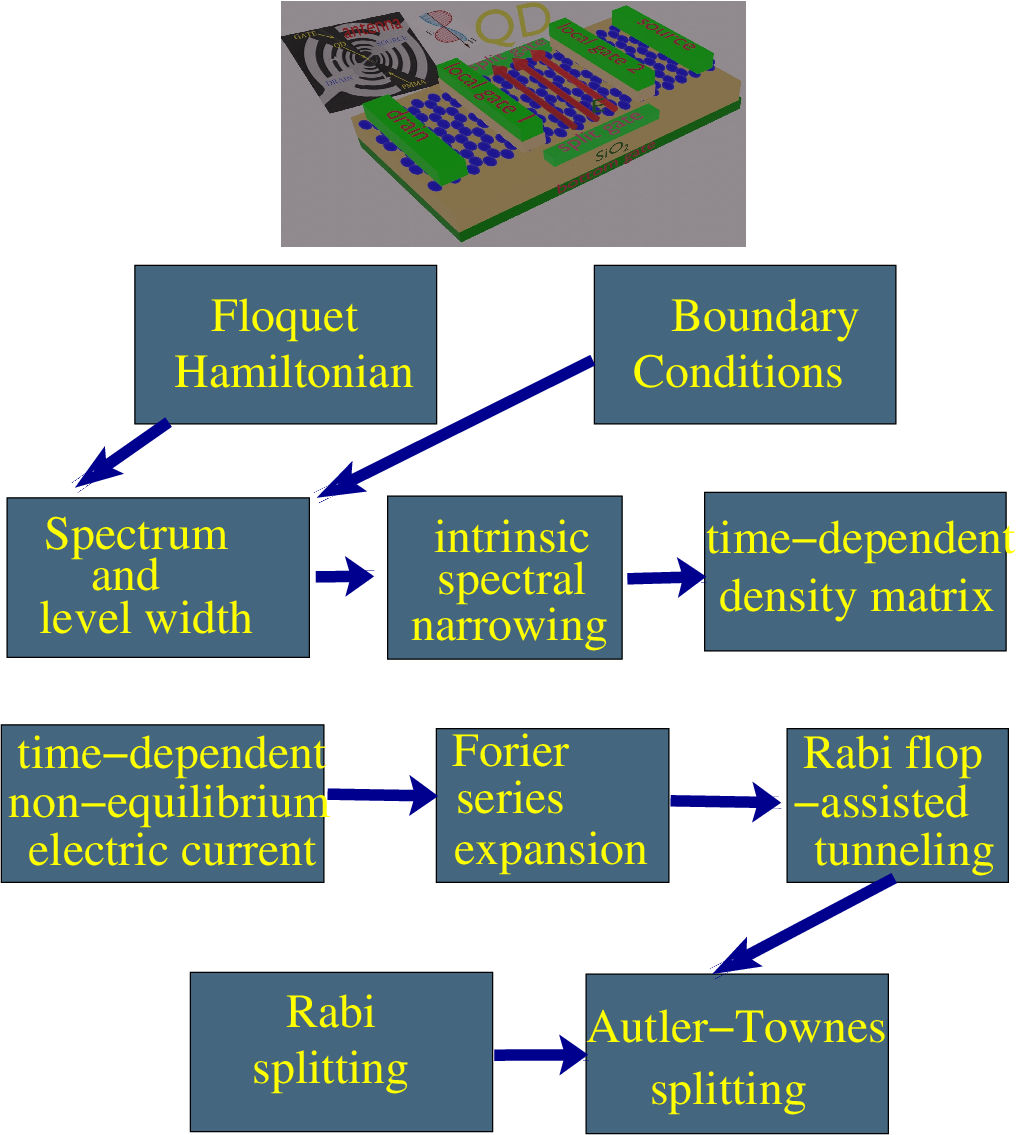} 
\caption{{The process flow diagram illustrating the approach.}}
\label{Fig_3}
\end{figure}

Let us consider the QD setup when the electric field vector $\mathbf{E}_{\mathrm{ac}} $ is directed perpendicular to the QD barriers in the piece-wise geometry shown in Figs.~\ref{Fig_1}, \ref{Fig_2}. The process flow diagram illustrating the approach is shown in Fig.~\ref{Fig_3}. We distiguish two different mechanisms of EF influence on QD. (i) The conventional photon-assisted tunneling (PAT)  \cite{Dayem-Martin,Tien-Gordon,Platero} under the influence of a time-dependent bias potential $V_{\rm bias} + V_{\mathrm{ac}}\cos \omega t$ represents the inelastic tunnel events when the electrons exchange energy quanta, i.e. photons, with the external EF. (ii) The Rabi flop-assisted tunneling (RFAT) caused by the time-cyclical behavior of TLS in the presence of EF. In TLS, the field induces cyclical absorption and re-emission of photons, whose time period depends on the detuning $\Delta = \omega -\omega _{Q}$, where $\omega $ is the EF frequency, $\omega_{Q}=\left( \varepsilon _{\alpha }-\varepsilon _{\beta }\right) /\hbar $ is the transition frequency related to the spacing (in the angular frequency units) between the quantized levels with energies $\varepsilon_{\alpha} $ and $\varepsilon_{\beta} $. 

In Section~\ref{sec3} we use the Floquet Hamiltonian (FH) \cite{Hirschfelder1,Hirschfelder2,Kirtman,Shavitt,Shirley,Breuer}, determine its eigenvalues,  eigenvectors, and characteristic exponents, as outlined in Sec.~A of Sec.~\ref{Supplement}. Then, using the solution obtained for the graphene stripe with zigzag edges  \cite{My-AQT,My-PRB23}, we apply the approach \cite{My-AQT,My-PRB23} to solve the boundary conditions for QD with the piece-wise geometry. The electron properties are described by Floquet formalism to compute how the electron eigenenergies and eigenvectors depend on the EF amplitude $E_{\mathrm{ac}}$ and frequency $\omega $. This also allows microscopic computing of the dipole matrix elements $\mathbf{d}=\left\langle i\right\vert e\mathbf{r}\cdot \mathbf{E}_{\mathrm{ac}}\left\vert j\right\rangle $ characterizing the EF-induced mixing of two quantum states $\left\vert i\right\rangle $ and $\left\vert j\right\rangle $ due to their coupling with each other in QD with the radius-vector $\mathbf{r}$. Besides, we calculate the EF-dependent electron transmission probability $T(\varepsilon )$. Derivation and computation of the PAT current are given in Sec.~C, D of Sec.~\ref{Supplement}. The modeling of RFAT is performed using the two QD models. (a) The simplest analytically solvable model assumes that QD contains just a single TLS, whose electron spectrum is unperturbed by EF. (b) In a more realistic model, we take into account the Autler--Townes effect \cite{AutTow} (the ac Stark effect) by including the Rabi splitting of quantized levels under the influence of EF.

By applying the almost degenerate perturbation formalism we find the matrix elements $\left\langle \alpha n\mid \lambda _{\beta m}\right\rangle $ of the Floquet Hamiltonian for TLS in the quantum dot resulting in analytical expressions for the density matrix $\rho _{\alpha \beta }\left( t\right) $. Then, in Subsection~\ref{sec4},  we calculate the non-stationary part $\delta n_{i}(t)$ of the electron distribution function $n_{i}$ related to Rabi flops. To find the electric current and the differential conductance in the voltage-biased QD we solve the Octavio-style boundary conditions \cite{Octavio1,Octavio2} (see Sec.~C,  in Sec.~\ref{Supplement} for details) for the electron distribution function $n_{i}$ and derive the respective non-stationary change $\delta n_{i}(t)$ of $n_{i}$ under the EF influence.
The obtained  $\delta n_{i}(t)$ and the QD transmission probability $T(\varepsilon )$ are then used in Subsections~\ref{sec3-3}, and \ref{sec3-6} to compute the electric current $I(V_{\mathrm{bias}}) $, the differential conductance $\sigma (V_{\mathrm{bias }})$, and Power Spectral Density of the voltage-biased QD, which is exposed to EF.

\section{TLS dynamics under the influence of ac field}\label{sec3}
\subsection{The density matrix of QD}\label{sec3-1}

The effect of interaction between QD and EF is schematically shown using an electric circuit in Fig. 2. We assume that QD is capacitively coupled to EF, whose effect is controlled electrically by change the positions $\varepsilon_{\alpha}$ and spacing $\omega _{\mathrm{Q}}$ of quantized levels. Technically, this is accomplished by applying the respective electric potentials to the local gate electrodes 1 and 2 as shown in Fig.~\ref{Fig_1} to change the height of potential barriers separating the sections of QD formed on the graphene stripe. On the one hand, the gate voltages control magnitudes of $\omega _{\mathrm{Q}}$ and $\varepsilon_{\alpha}$. On the other hand, EF induces the dipole coupling between the quantized states resulting in Rabi flops that are accompanied by cyclical absorption and re-emission of photons and by the oscillations of the population of the quantized levels. 
In this way, one accomplishes the system where $\varepsilon_{\alpha}$ and  $\omega _{\mathrm{Q}}$ are controlled by the dc voltages, while the interaction between the quantized states is controlled by EF. Essentially, the parameters of quantized states as well as dynamics and features of the EF-induced interstate interaction are encoded in the Rabi flops magnitude and shape of their time dependence. This paper suggests how to determine those parameters directly by measuring the I-V curves of QD.

We first consider the simplest model of the two-level system (TLS) located inside QD and exposed to the external EF \cite{Hirschfelder1,Hirschfelder2,Kirtman,Shavitt,Shirley,Breuer}. In the model with a single TLS, many important characteristics are computed analytically allowing a deeper qualitative understanding of the dipole coupling between the quantized level system and a semiclassical or quantized EF. Below we use analytical expressions for the density matrix (see Secs.~A and B in Sec.~\ref{Supplement}) and electron transmission probability through QD, which depend on the frequency $\omega $ and amplitude $\mathbf{E}_{\mathrm{ac}}$ of the exciting ac field (see Secs.~C and D in Sec.~\ref{Supplement}). The TLS dynamic under the influence of a semiclassical field is described by a time-periodic Hamiltonian~\cite%
{Hirschfelder1,Hirschfelder2,Kirtman,Shavitt,Shirley,Breuer} as described in (see Secs.~A and B in Sec.~\ref{Supplement}). Using the Floquet formalism, we transform the time-dependent Hamiltonian of a single TLS into the time-independent Floquet Hamiltonian (FH) $\mathcal{H}_{\mathrm{F}} $ whose eigenvalues and eigenvectors serve to construct solutions of the original wavefunction equation as outlined in Sec.~A of Sec.~\ref{Supplement}. The diagonal elements of the infinite-dimensional  $\mathcal{H}_{\mathrm{F}} $ become degenerate in pairs in the vicinity of a resonant transition  (either single-photon or multiphoton) between the TLS levels. Technically, $\mathcal{H}_{\mathrm{F}} $ is diagonalized using the almost degenerate perturbation theory \cite{Hirschfelder1,Hirschfelder2,Shirley} allowing us to compute the energy eigenvalues with no need to calculate very high-order wave functions. The most interesting case relates to the two-fold nearly degenerate eigenvalues when neglecting all the other states \cite{Hirschfelder1,Hirschfelder2,Shirley}, which actually corresponds to the rotating field approximation.

\begin{figure}
\includegraphics[width=85 mm]{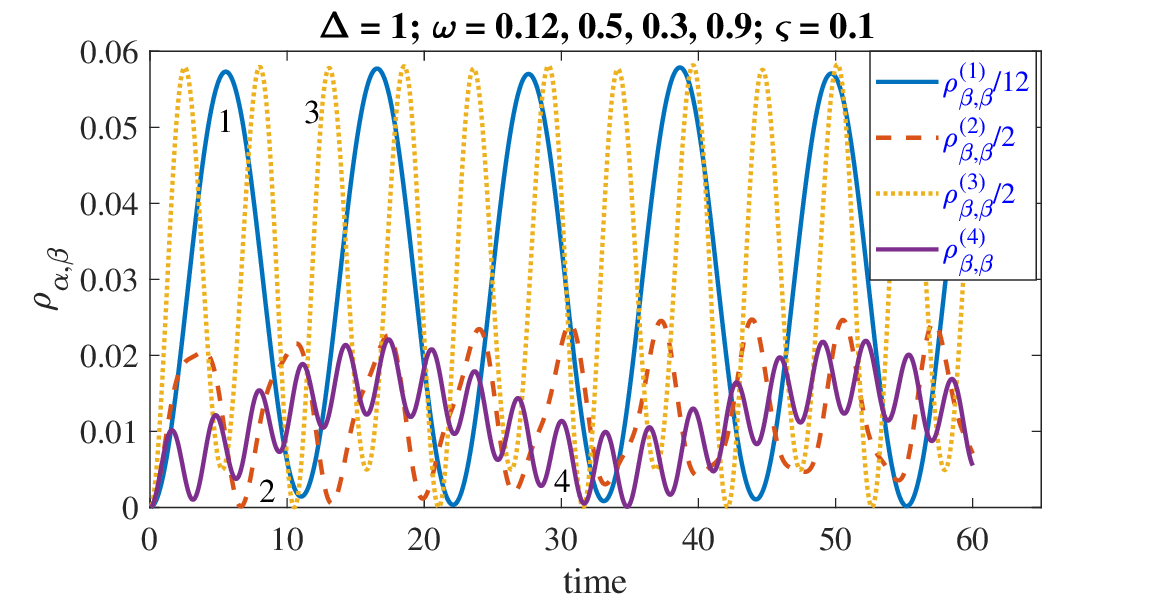} 
\caption{ The time dependence of diagonal part of the density matrix element $\rho _{\alpha \beta }\left( t\right) $ of the lower state ($\alpha = 1$, $\beta = 1$) averaged over initial times near the primary resonance $\omega $ (see Ref.~\cite{Hirschfelder1}), for $\varsigma =0.1$, $\Delta =1$, $ \omega =0.12$, $0.5$, $0.3$, and $0.9$ (see curves 1-4 respectively). The Floquet state $\left\vert \alpha 0\right\rangle $ and $\left\vert \beta 0\right\rangle $ are nearly degenerate, while all other states are relatively far away.
}
\label{Fig_4}
\end{figure}

\begin{figure}
\includegraphics[width=85 mm]{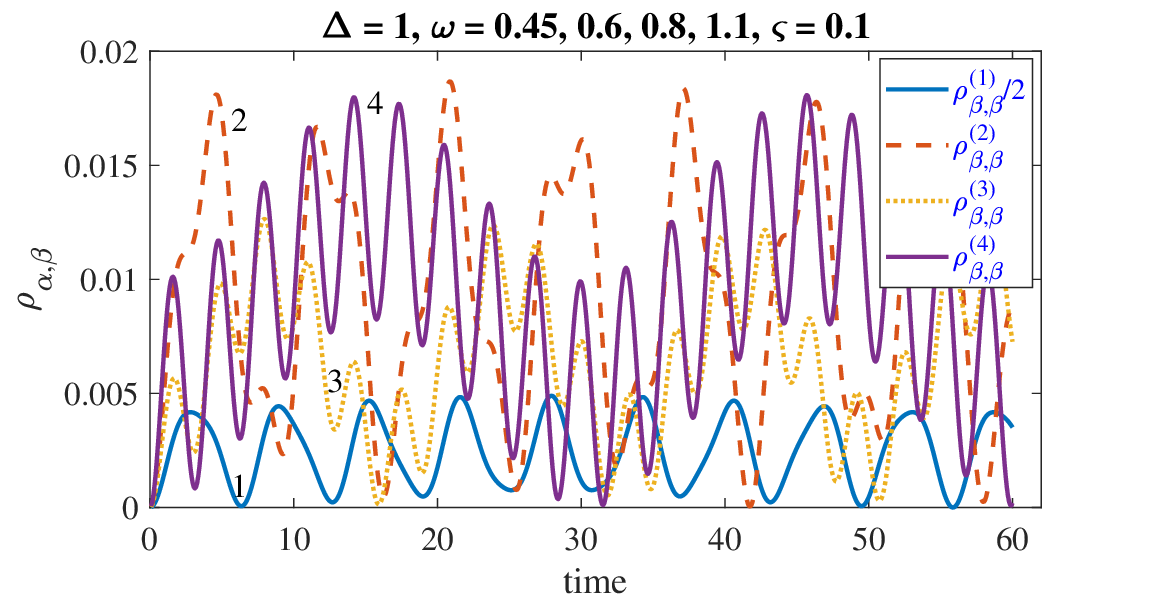} 
\caption{ {Rabi flops in the density matrix averaged over initial times near the primary resonance $\omega $, according to Eq.~(\ref{rho-ab}) for $\varsigma =0.1$, $\Delta =1$, $\omega =0.45$, $0.6$, $0.8$, and $1.1$ (see curves 1-4 respectively).
}}
\label{Fig_5}
\end{figure}

\begin{figure}
\includegraphics[width=85 mm]{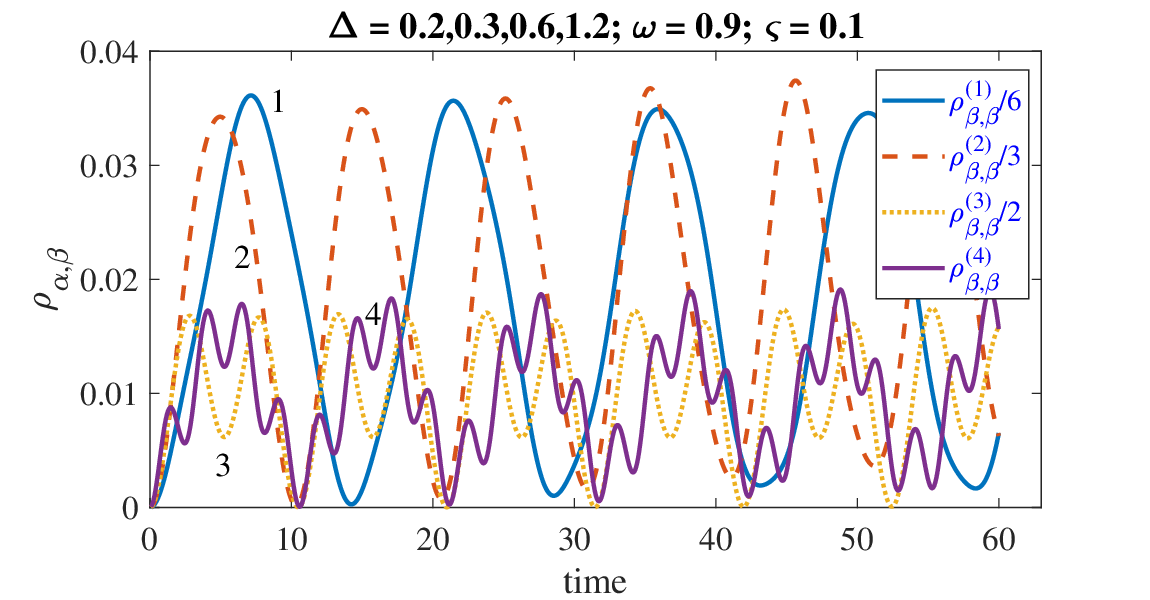} 
\caption{ { The density matrix averaged over initial times near the primary resonance for $\varsigma =0.1$,  $\omega =0.9$, and $\Delta =0.2$, $0.3$, $0.6$, and $1.2$, (see curves 1-4 respectively).}}
\label{Fig_6}
\end{figure}

\begin{figure}
\includegraphics[width=85 mm]{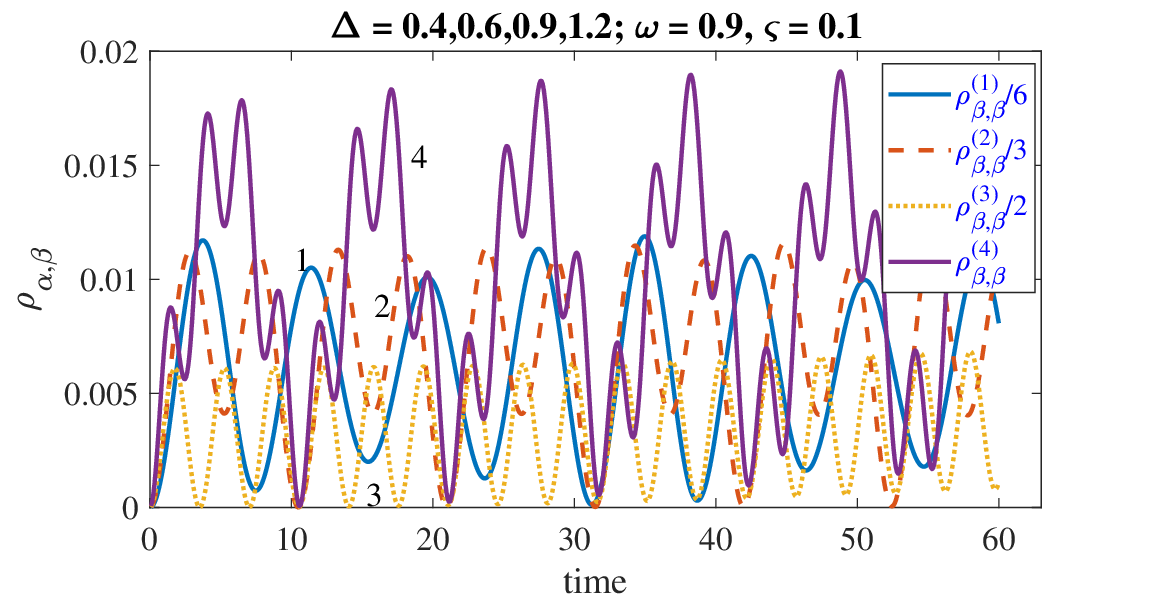} 
\caption{ {The density matrix averaged over initial times near the primary resonance for $\varsigma =0.1$, $\Delta =0.4$, $0.6$, and $1.2$, $\omega = 0.9$  (see curves 1-3 respectively).}}
\label{Fig_7}
\end{figure}

\begin{figure}
\includegraphics[width=85 mm]{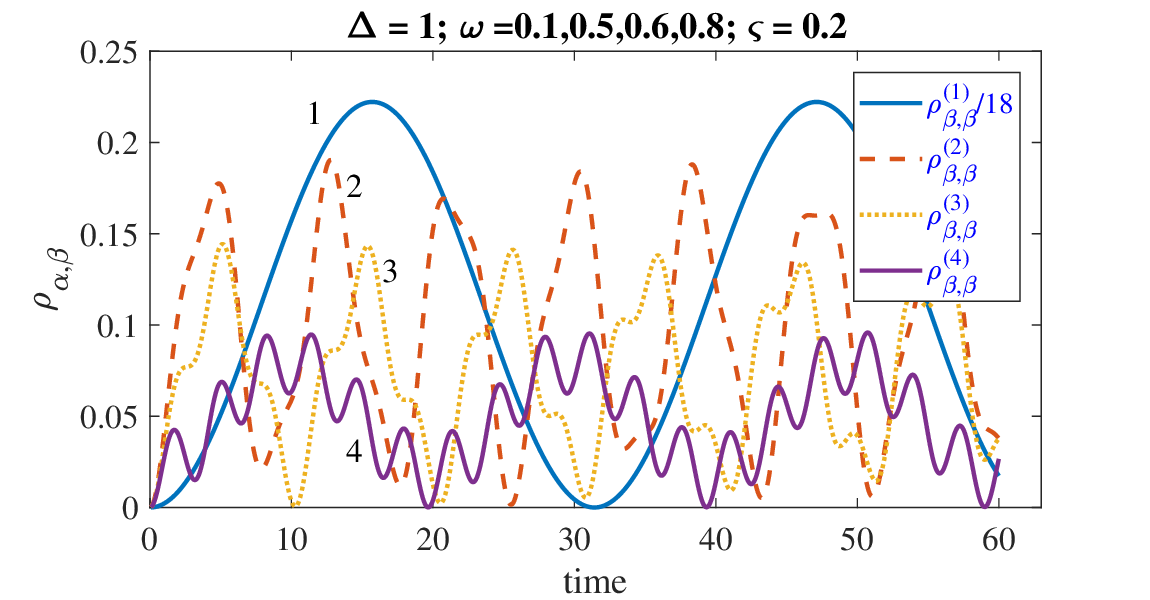} 
\caption{ {The density matrix averaged over initial times near the primary resonance  computed for $\varsigma =0.2$. The EF frequency values for density matrices  $\rho^{(1-4)}_{\beta,\beta}$ respectively are $\omega =0.1$, $0.5$, $0.6$, and $0.8$.}}
\label{Fig_8}
\end{figure}

\begin{figure}
\includegraphics[width=85 mm]{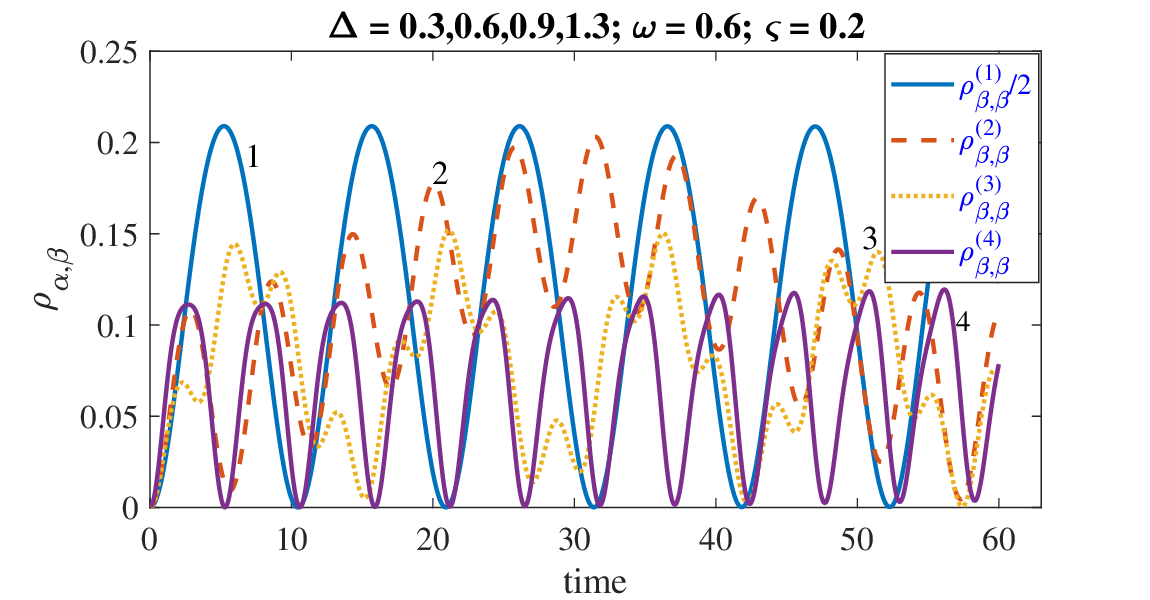} 
\caption{ { The density matrix averaged over initial times near the primary EF resonance
with frequency $\omega =0.6$, for $\varsigma =0.2$ computet from Eq.~(\ref{rho-ab}). The
detuning $\Delta $ values for density matrices  $\rho^{(1-4)}_{1,1}$ respectively are $0.3$, $0.6$, $0.9$, and $1.3 $.}}
\label{Fig_9}
\end{figure}

Following the Floquet approach \cite%
{Breuer,VanVleck,Kemble,HaugKoch,Levi-book,Blum-book,Cohen-T} (see Sec.~A of Sec.~\ref{Supplement}) in the dipole approximation \cite{Hirschfelder1,Hirschfelder2,Shirley}, the diagonal part of the density matrix $\rho _{\beta \beta }\left(t\right) $ in the central section C of QD averaged over initial times is obtained in the form%
\begin{eqnarray}
\rho _{\beta \beta }\left( t\right)  &=&\frac{\varsigma ^{2}}{\omega_{\rm R}^{\prime 2}}%
\left( 1-\frac{\varsigma ^{2}}{\omega ^{2}}\right) ^{2}\sin ^{2}\varpi_{\rm R} t+%
\frac{\varsigma ^{2}}{2\omega ^{2}}  \nonumber \\
&&-\frac{\varsigma ^{2}}{4\omega ^{2}}[\frac{1}{\omega_{\rm R}^{\prime 2}} \left( \varsigma ^{2}+\left(\omega_{\rm R}^{\prime 2}+\Delta ^{\prime 2}\right) \cos 2\varpi_{\rm R} t\right) \cos
2\omega t  \nonumber \\
&&+2\frac{\Delta ^{\prime }}{\omega_{\rm R}^{\prime } }\sin 2\varpi_{\rm R} t\sin 2\omega t]
\label{rho-ab}
\end{eqnarray}%
where we introduced the renormalized Rabi frequency $\omega_{\rm R}^{\prime} =\sqrt{\varsigma ^{2}+\Delta ^{\prime 2}}$. Here $\varsigma $ is the interlevel coupling strength,
\begin{equation}
\varpi_{\rm R}^{2}=\varsigma ^{2}\left( 1-\frac{\varsigma ^{2}}{4\omega ^{2}}%
\right) ^{2}+\left( \Delta -\frac{\varsigma ^{2}}{2\omega }\right)
^{2}\left( 1-\frac{\varsigma ^{2}}{2\omega ^{2}}\right) ^{2}+\mathcal{O}%
\left(\frac{\varsigma ^{5}}{\omega ^{3}}\right)\text{,}
\end{equation}%
or approximately
\begin{equation}
\varpi_{\rm R} \approx \left( 1-\frac{\varsigma ^{2}}{4\omega ^{2}}\right) \omega_{\rm R}^{\prime} \text{,}
\end{equation}%
the renormalized detuning is
\begin{equation}
\Delta ^{\prime }=\Delta -\frac{\varsigma ^{2}}{2\omega }-\frac{\Delta \varsigma ^{2}}{4\omega ^{2}}\text{,}
\end{equation}%
and the detunning is
\begin{equation}
\Delta =\omega -\omega _{Q}   \text{,}
\end{equation}%
where $\varsigma $ is the dipole coupling strength between the electron states $\alpha $ and $\beta $, $\varpi_{\rm R} $ is the modified $\omega_{\rm R}^{\prime }$, $\omega $ is the external field frequency, $\omega _{Q}=\left( \varepsilon _{\alpha }-\varepsilon _{\beta }\right) /\hbar $ is the energy level separation in the angular frequency units. We emphasize that $\rho _{\beta \beta }\left( t\right) $ is also related to the probability of transition $P_{\beta \leftarrow \alpha }\left( t\right) $ between $\alpha $ and $\beta $ states, so $P_{\beta \leftarrow \alpha }\left( t\right) =\rho _{\beta \beta }\left( t\right) $. For the sake of simplicity, in Eq. (\ref{rho-ab}) we omitted the initial-time dependence, which includes many extra terms, whose influence is not observed  experimentally. The time dependence of $\rho _{\alpha \beta }\left( t\right) $ is illustrated in Figs.~\ref{Fig_4}, \ref{Fig_5}, \ref{Fig_6}, \ref{Fig_7}, \ref{Fig_8}, and \ref{Fig_9} for different values of $\varsigma $, $\Delta $, and $\omega $. In the mentioned figures we plot the time dependence of the density matrix component for the lower level, $\alpha = \beta =1$. The time is in units of $1/(\hbar \omega_{\rm R})$, $\varsigma$, $\Delta$, and $\omega $ are in units of the energy level spacing $\varepsilon_{\alpha } - \varepsilon_{\beta }$.

We also determine the pseudoenergies $W_k$, directly by using a digital computer
with a standard algorithm to solve the finite-order secular equations which
correspond to truncating Floquet Hamiltonian. In principle, this permits
finding the solution to any desired precision. The exact results were
obtained by truncating $\mathcal{H}_{\mathrm{F}}$ and using a standard
computer algorithm to determine its eigenvalues directly.

\subsection{Rabi-type oscillation in the population of the quantized levels}\label{sec4}

The Fermi-Dirac electron distribution function \cite{Breuer,Kemble,Blum-book} 
$n_{i}^{\left( 0\right) } $ is computed using the steady-state density matrix $\rho ^{\left( 0\right) }$ [see Eq.~(\ref{rho-ab})] as 
\begin{equation}
 n_{i}^{\left( 0\right) } =\mathrm{Tr}\left[ \rho
^{\left( 0\right) }N_{i}\right] =1/\left( \exp \left( \left( \varepsilon
_{i}-\mu \right) /k_{\mathrm{B}}T\right) +1\right)  \label{n0}
\end{equation}%
where $\mu $ is the electrochemial potential, $T$ is the temperature, $k_{%
\mathrm{B}}$ is the Boltzmann constant. The Rabi flops cause the
time-dependent oscillatory change $\delta n_{i}(t) $ in the distribution function $%
n_{i} = n_{i}^{\left( 0\right) }  +  \delta n_{i}(t) $ related to the single-particle energy eigenstate $\varepsilon _{i}$ \ localized in QD. One gets
\begin{equation}
\delta n_{i}\left( t\right) =\mathrm{Tr}\left[ \delta \rho \left( t\right)
N_{i}\right] \text{,}  \label{dnt}
\end{equation}%
where $N_{i}$ is an observable occupation number of the $i$-th
single-particle energy state $N_{i}\left\vert \mathbf{n}\right\rangle
 = n_{i}\left\vert \mathbf{n}\right\rangle $. The non-stationary addition $%
\delta n_{i}\left( t\right) $ is computed using either the von Neumann equation or the non-equilibrium technique \cite%
{Keldysh1964}. In this paper we use Floquet formalism to compute $n_{i} $
\begin{equation}
n_{i}=\mathrm{Tr}\left( \rho N_{i}\right) =\mathrm{Tr}\left[ \left( \rho
^{\left( 0\right) }+\delta \rho \left( t\right) \right) N_{i}\right]
=n_{i}^{\left( 0\right) }+\delta n_{i}\left( t\right)  \label{ni}
\end{equation}

\subsection{Photon-assisted tunneling through QD}\label{sec3-3}

Let us consider QD exposed to EF, whose effect is depicted using the electric circuit in Fig.~\ref{Fig_2}.The derivation of the PAT current through the quantum dot is given in Secs.~C, D of Sec.~\ref{Supplement}l. We use the Octavio-style boundary conditions \cite{Octavio1,Landauer,Buttiker,Octavio2}, which are valid for arbitrary transparency of QD interfaces \cite{Datta-1995} and take into account the non-stationary electron distribution inside QD. The time dependence of $\delta n_{i}\left( t\right) $ originates from Rabi flops considered in the former Sec.~\ref{sec3-1}. Then, according to Eq.~(S30) derived in Sec.~\ref{Supplement}, the electric current $I_{\rm QD}$ flowing through QD, becomes time-dependent as well resulting in the RFAT ladder formed in the IV curve. During the numeric computing of the RFAT ladder, we use analytical expressions for the density matrix $\rho _{\alpha \beta }\left( t\right)$ provided in Sec.~D. We also use the time-dependent electron distribution function $n_{i}\left( t\right)$ given by Eq.~ (\ref{ni}) in the central quantum dot section C, which allows computing the respective electric current. These allow examining the effect of Rabi flops, which are expressed in the time dependence of $\delta n_{i}\left( t\right)$ and electric current.

\begin{figure}
\includegraphics[width=85 mm]{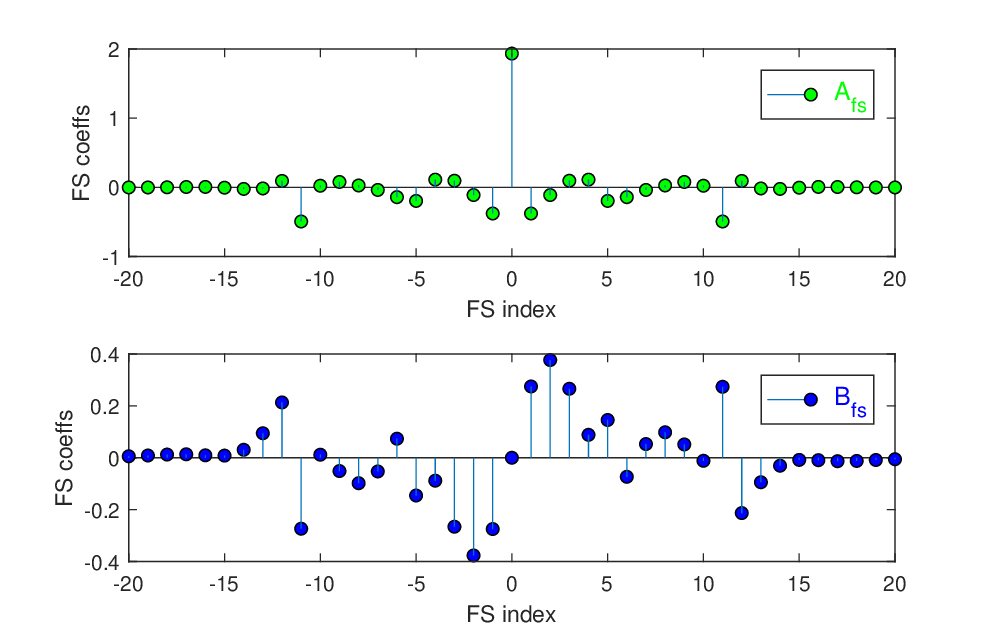} 
\caption{ {The Fourier coefficients $A_{n}$ and $B_{n}$ in expansion of the electric current $I_{\rm QD}(t) $ vs the Rabi flop harmonic index $n$ in QD.}}
\label{Fig_10}
\end{figure}

The schematics of the device where the gate voltages $V_{\rm lg}$ are applied to the local gate electrodes of QD to control the heights of potential barriers separating the sections of QD formed on the graphene stripe is illustrated in Figs.~\ref{Fig_1}, \ref{Fig_2}. We consider the effect of external EF acting on the local gates. The ac field induces the ac voltage drop $V_{\mathrm{ac} }\cos \omega t$ between the central and adjacent sections of the quantum dot. Here we neglect the interaction of the ac field with the barrier and consider one of the adjacent sections as a reference (e.g., the left region). The microwave field causes an additional voltage drop $V_{\mathrm{ac}}\cos \omega t$ on the top of the dc voltage $V_{\rm bias}$ in the central section. Within this simple model, the effect of the external field is accounted for by adding a time-dependent, but spatially homogeneous potential within the central section which is described by a local Hamiltonian%
\begin{equation}
\mathcal{H}=\mathcal{H}_{0}+eV_{\mathrm{ac}}\cos \omega t
\end{equation}%
The time-dependent homogeneous potential does not affect the spatial distribution of the electronic wave function within each region. In Sec.~D of Sec.~\ref{Supplement}, we derive the expression (S17) for the time-dependent electric current $I_{\rm QD}\left( t\right)$ as expansion with the trigonometric coefficients $A_{0}$, $A_{n}$, and $B_{n}$ (see plots in Fig.~\ref{Fig_10}). We also compute the trigonometric Fourier series expansion of the time-dependent electric current $I_{\rm QD} (t)$ in QD. These greatly simplify numeric calculations of the conventional PAT and RFAT currents.

To better understand how the different microscopic mechanisms involved in PAT show up in the electric current flowing through QD, in our calculations, we simply assume that only one certain mechanism acts at a time. Practically, this helps sort out and identify the relevant mechanisms when measuring the PAT IV curves experimentally.

\begin{figure}
\includegraphics[width=85 mm]{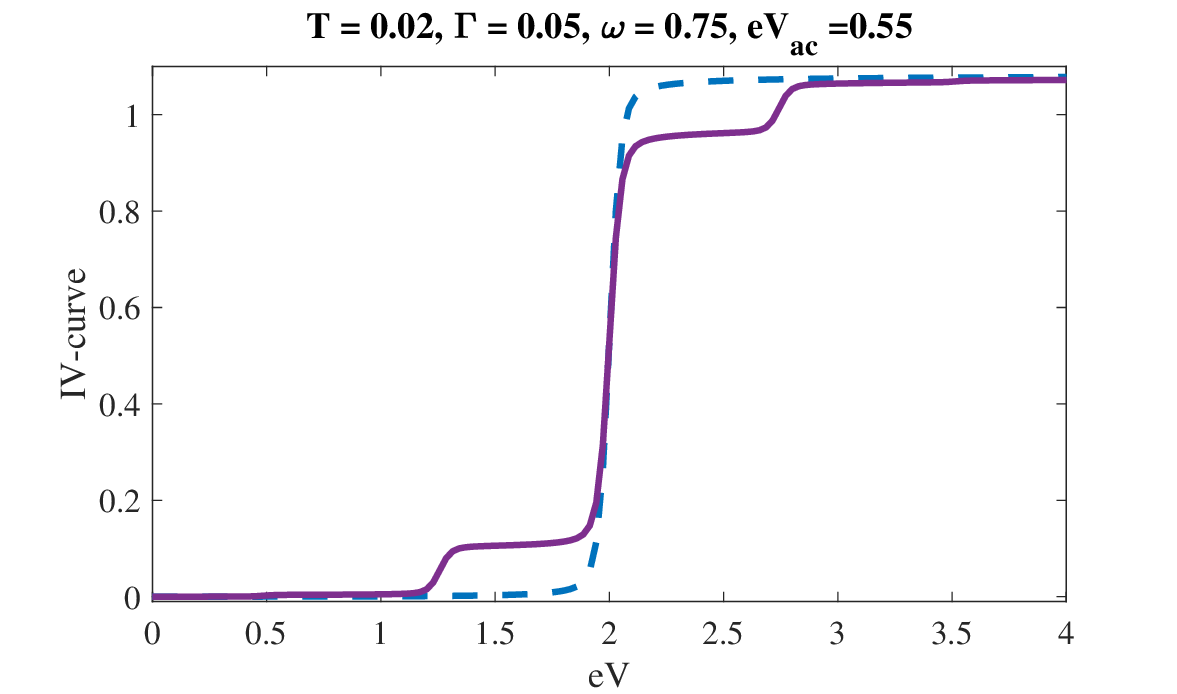} 
\caption{ {The conventional photon-assisted tunneling \cite{Dayem-Martin,Tien-Gordon,Platero} (PAT) through QD. Here the dashed blue curve is for the steady state and the pink curve is the PAT ladder induced by EF.}}
\label{Fig_11}
\end{figure}

Let us first briefly discuss the conventional PAT when the mixed quantum state in QD for some reason is not formed because it is destroyed, e.g., by noise or by temperature fluctuations. It was the case in the experimental measurement of the differential conductance of the CNT quantum dot exposed to the external THz irradiation reported in Ref.~\cite{Kawano1,Rinzan,Kawano2}. Here we do not consider the Coulomb blockade effect that was studied before in detail by many researchers (see, e.g., Ref.~\cite{Kouwenhov} and references therein).
They observed pronounced change in the junction$^{\prime }$s conductance due to the photon-assisted tunneling (PAT) of single electrons \cite{Kawano1,Rinzan,Kawano2} regarded as the Dayem-Martin effect \cite{Dayem-Martin,Tien-Gordon,Platero}.
Assuming the absence of Rabi flops, the calculation gives the conventional PAT ladder shown in Fig.~\ref{Fig_11}, where $T = 0.02$, $\Gamma =0.05$, $\omega = 0.75$, and $eV_{\rm ac} = 0.55$ (all are given in units of $\varepsilon_{\alpha } - \varepsilon_{\beta }$). In this system of units, we set $k_{\rm B} =1 $, $e = 1$, and $\hbar =1$. Then, e.g, if we use $\omega = 7.5$~THz, we get $T = 10$~K, $\Gamma = 2$~meV, the energy relaxation time $\tau_{\varepsilon } = 3\times 10^{-13} $~s, and $eV_{\rm ac} = 23$~meV.  In Fig.~\ref{Fig_11} one observes the Dayem-Martin steps spaced by $\Delta V_{\mathsf{DM}}=\hbar \omega/e $ originating from PAT as was reported in Refs.~\cite{Kawano1,Rinzan,Kawano2,Brune}.

In the next step, we consider RFAT, which is not related to the absorption/emission of the external EF photons with energy $\hbar \omega $ but rather involves the other photons with Rabi energy  $\hbar \omega_{\rm R} $, which are absorbed/emitted in the course of the Rabi flops. To better understand how RFAT is pronounced in the IV curves, below we disregard the conventional PAT~ \cite{Dayem-Martin,Tien-Gordon,Platero} and Coulomb blockade ~\cite{Kouwenhov} while focusing our attention exclusively on RFAT.

\begin{figure}
\includegraphics[width=85 mm]{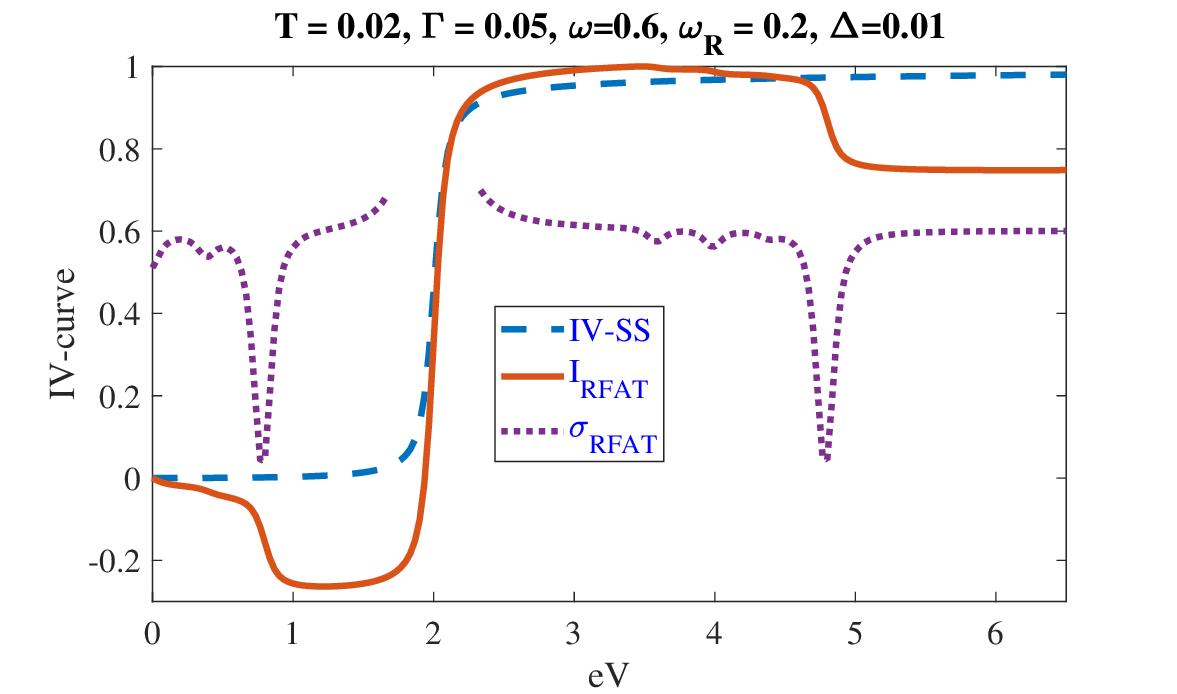} 
\caption{{The I-V curve of QD, which is modeled as the double barrier NICIN junction irradiated by EF with frequency $\omega = 0.6$. The temperature is $T = 0.02$, the energy level width $\Gamma = 0.05$, Rabi frequency $\omega_{\rm R} = 0.2$, and detunning $\Delta = 0.01$. One can distinguish descending steps spaced by $\Delta V_{\mathsf{bias}} = 0.4 \simeq 2\hbar \omega_{\rm R}$. To better see the effect of Rabi flop-assisted tunneling (RFAT), we also plot the differential conductance $\sigma_{\rm RFAT} (V)$ (shown here and below in arbitrary units), whose dips (or peaks) emphasize the Rabi flops on the I-V curve.}}
\label{Fig_12}
\end{figure}

\begin{figure}
\includegraphics[width=85 mm]{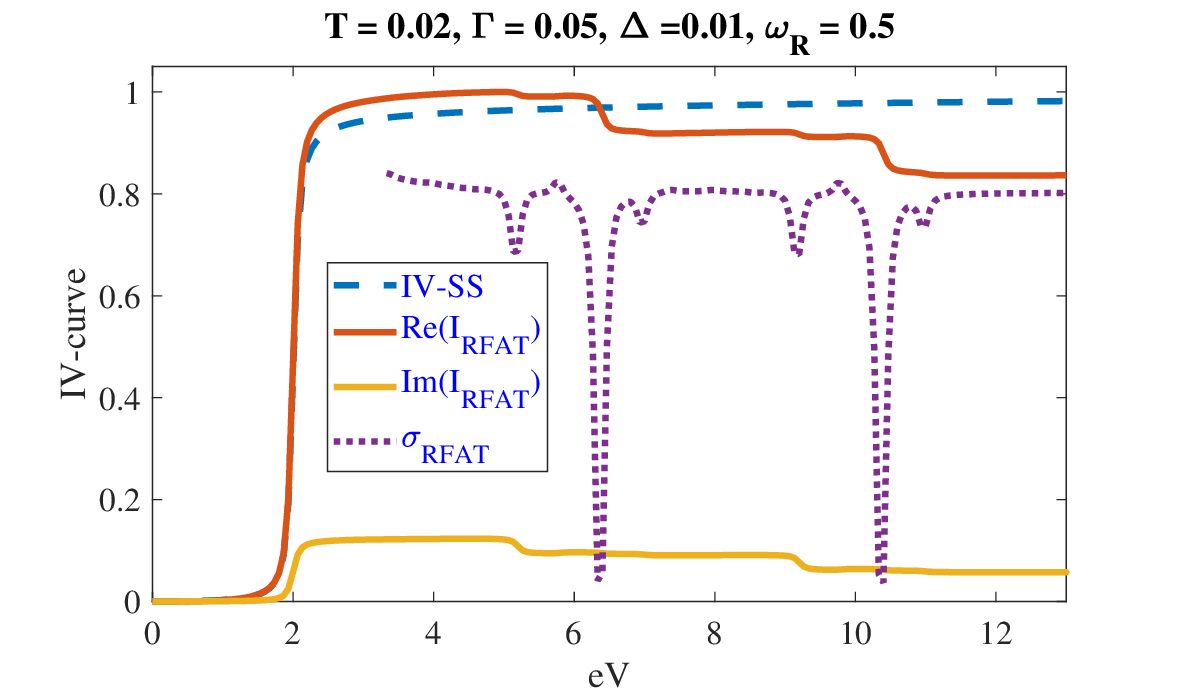} 
\caption{{The Rabi flop-assisted tunneling through QD. The Rabi flop contribution in the I-V curve has the form of descending steps reducing the electric current downward and spaced by $\omega _{\mathrm{R}} = 0.5$. We also show the differential conductance $\sigma _{\mathrm{RFAT}}\left( V_{\mathrm{bias}}\right) $, whose respective extremums are spaced by $\Delta V_{\mathsf{bias}}=\hbar \omega _{\mathrm{R}}/e=0.5$  corresponding to the photon absorption in the successive Rabi flops. Two positive weak peaks are visible at $eV \simeq 5.5$ and $ \simeq 9.5$ denoting weak photon emission. The respective dips of differential conductance $\sigma _{\mathrm{RFAT}}\left( V_{\mathrm{bias}}\right) $ emphasize the RFAT steps.}}
\label{Fig_13}
\end{figure}

\begin{figure}
\includegraphics[width=85 mm]{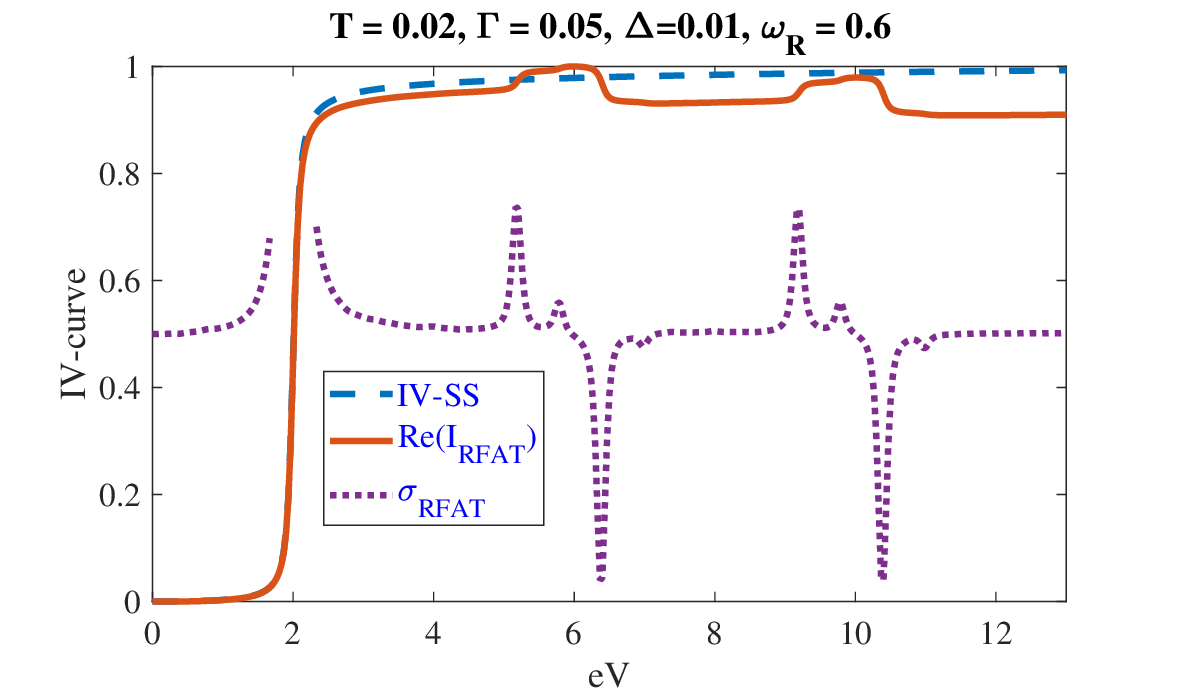}
\caption{ {The Rabi flop-assisted tunneling through QD. The detailed structure of the above RFAT I-V curve is reflected in the $\sigma _{\mathrm{RFAT}}\left( eV\right) $ curve. In this $\sigma _{\mathrm{RFAT}}\left( eV\right) $ curve along with sharp dips corresponding to the one-photon emission, there are also peaks related to the one-photon absorption.}}
\label{Fig_14}
\end{figure}

\begin{figure}
\includegraphics[width=85 mm]{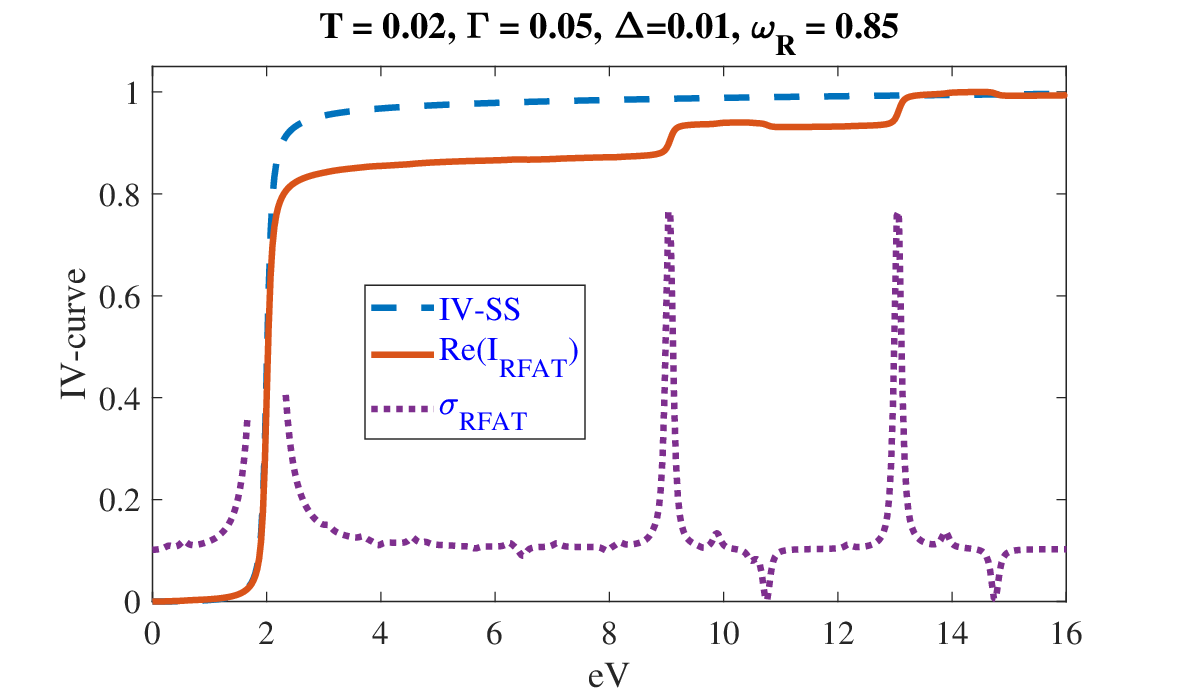} 
\caption{ { The RFAT I-V curve shows a series of descending and ascending steps spaced by $\Delta V_{%
\mathsf{bias}}=2\hbar \omega \Omega _{\mathrm{R}}/e=0.85$. Such steps
originate from respective decreases or increases of the RFAT electric
current accompanied by the one-photon absorption processes. The one-photon
absorption prevails when $\omega _{\mathrm{R}}$ increases to $\omega _{%
\mathrm{R}}=0.85$.}}
\label{Fig_15}
\end{figure}

\begin{figure}
\includegraphics[width=85 mm]{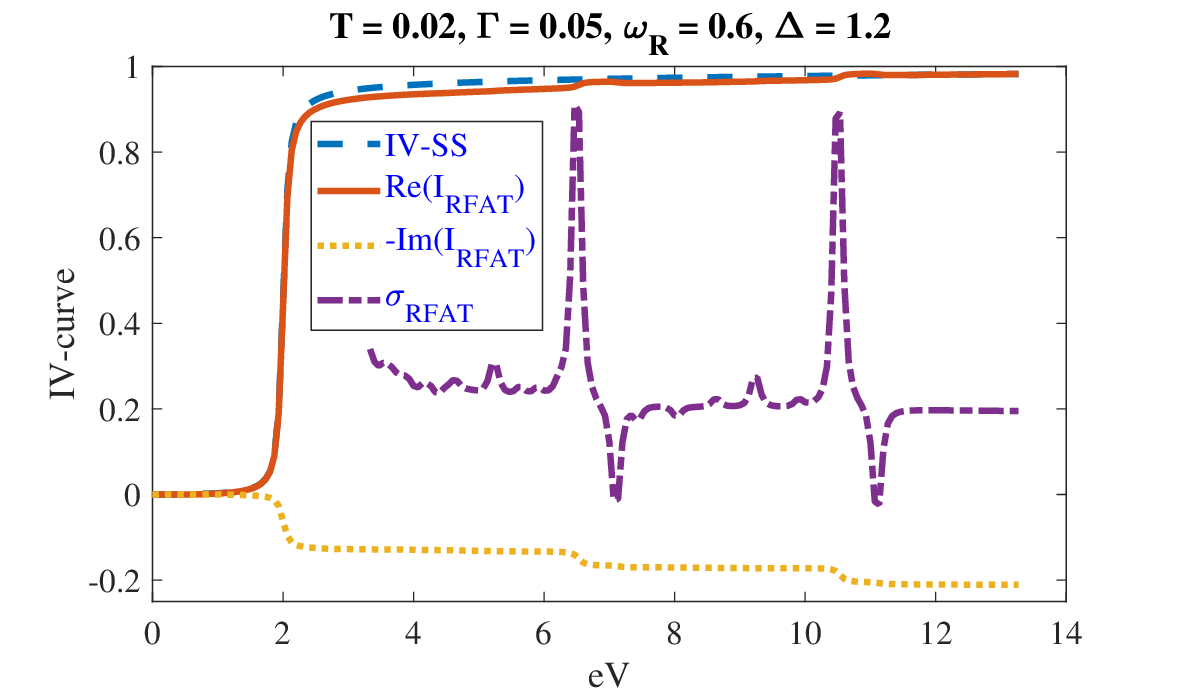} 
\caption{{For larger $\omega _{\mathrm{R}}$, the photon emission becomes so strong that it redirects the reactive part of electric current  (i.e., $\Im m(I_{\rm RFAT})$) to the opposite.}}
\label{Fig_16}
\end{figure}

\begin{figure}
\includegraphics[width=85 mm]{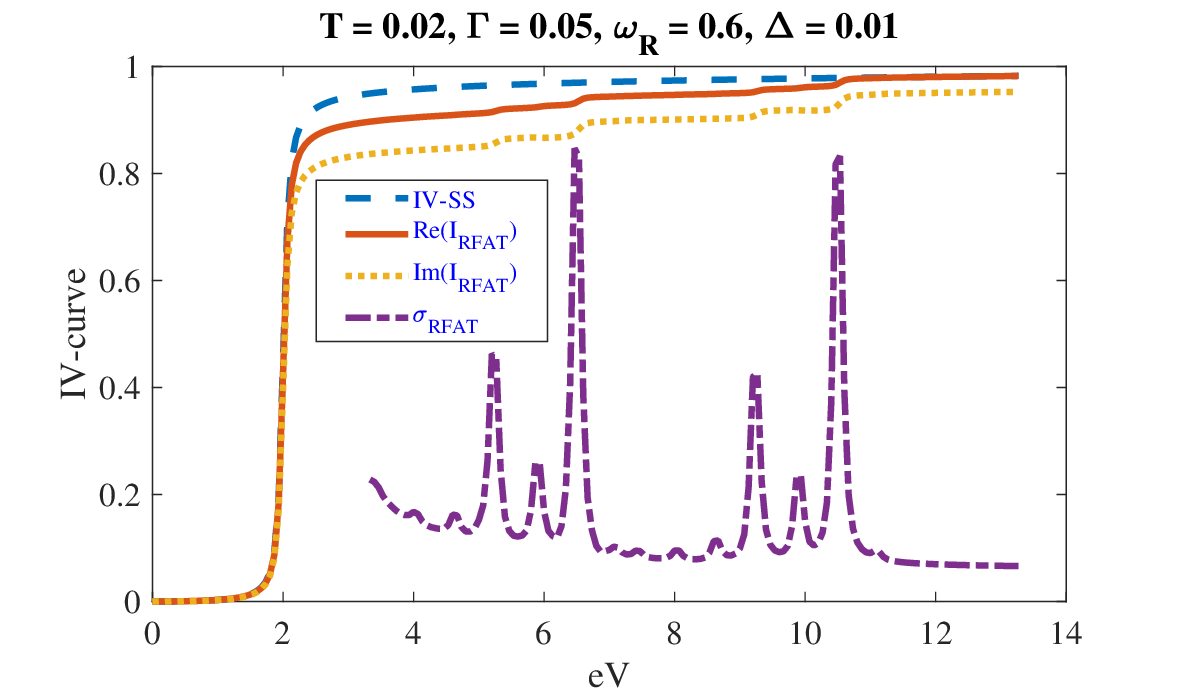} 
\caption{{At certain combinations of QD and EF parameters, all the RFAT steps become ascending, which corresponds to the EF-stimulated photon absorption.}}
\label{Fig_17}
\end{figure}

\begin{figure}
\includegraphics[width=85 mm]{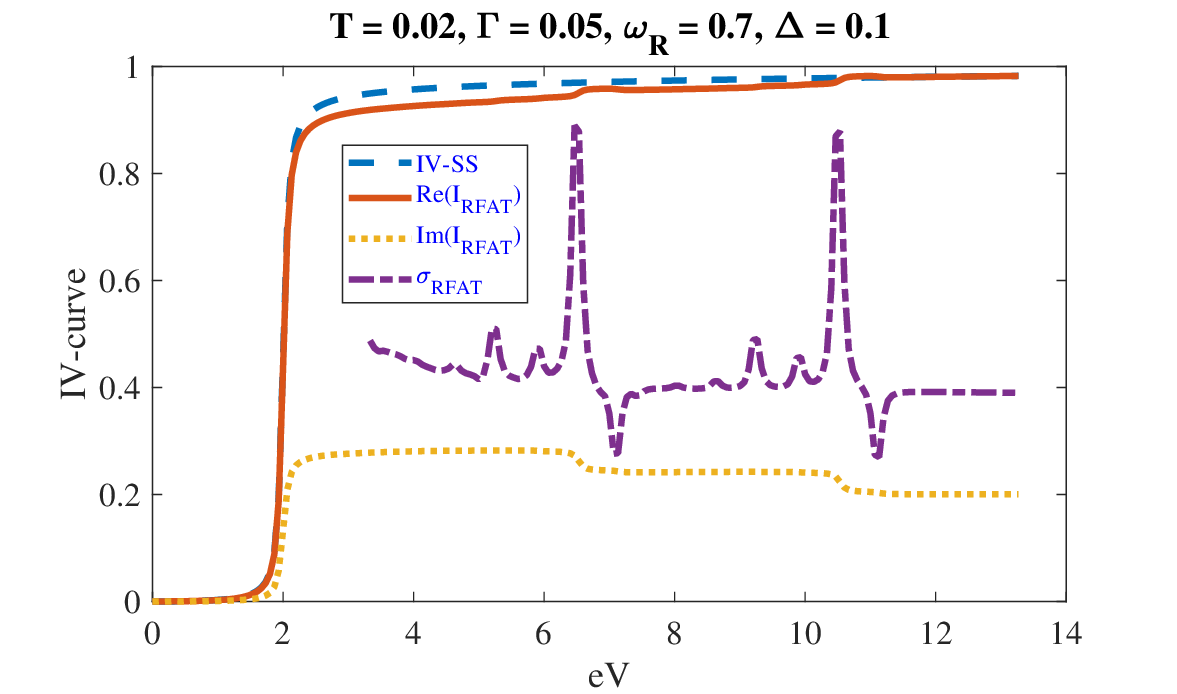} 
\caption{ {Here we also show the imaginary part of the RFAT current (dotted curve)}}
\label{Fig_18}
\end{figure}

The calculation results for RFAT in QD with a single TLS are presented in Figs.~\ref{Fig_12}--\ref{Fig_18}, where we show the RFAT I-V curves computed for different values of temperature $T$, level width $\Gamma $, detuning $\Delta $, energy level spacing $\varepsilon _{\alpha }-\varepsilon _{\beta } $, and Rabi frequency $\omega _{\mathrm{R}}$. One can see that the obtained RFAT I-V curves show remarkable series of descending and ascending steps spaced by $\Delta V_{\rm{bias}}=n\hbar \omega _{\mathrm{R}}/e$ ($n$ is integer) originating from respective decreases or increases of the RFAT electric current accompanied by the one-photon absorption processes. 

Again, in Figs.~\ref{Fig_12}--\ref{Fig_18} we use the same units as in the former Fig.~\ref{Fig_11}. They are translated to the conventional untis as follows. If in Fig.~\ref{Fig_13} we use $\omega = 7.5$~THz, our units there would correspond to $T = 10$~K, $\tau_{\varepsilon } = 3\times 10^{-13} $~s, $\omega_{\rm R} = 1.3\times 10^{13}$~s$^{-1}$, and $\Delta = 6.3\times 10^{11}$~s$^{-1}$.

\subsection{Power Spectral Density}\label{sec3-6}

The power spectrum $S(\omega )$ of a time series $I_{\mathrm{QD}}\left(
t\right) $ describes the distribution of power into frequency components
composing the electron current in QD. According to Fourier analysis, $I_{%
\mathrm{QD}}\left( t\right) $ can be decomposed into a number of discrete
frequencies, or a spectrum of frequencies over a continuous range. The
statistical average of a certain $I_{\mathrm{QD}}\left( t\right) $
(including noise) as analyzed in terms of its frequency content, is called
its spectrum.

Next we compute the energy spectral density by assumig that the energy of the signal is concentrated around a finite time interval, provided its total energy is finite. This serves as a good approximation when the signals exist over all time, or over a time period large enough (in respect to the duration of a measurement).
The power spectral density (PSD) then refers to the spectral energy
distribution that would be found per unit time, since the total energy of
such a signal over all time would generally be infinite. Summation or
integration of the spectral components yields the total power (for a
physical process) or variance (in a statistical process), identical to what
would be obtained by integrating $I_{\mathrm{QD}}^{2}(t)/R_{\mathrm{QD}}$
over the time domain ($R_{\mathrm{QD}}$ is the QD dc resistance), as
dictated by Parseval's theorem \cite{Parseval}.

The spectrum of the non-stationary current $I_{\mathrm{QD}}\left( t\right) $
in QD contains essential information about the nature of $I_{\mathrm{QD}}$.
For instance, the Rabi flop dynamics similarly to the pitch and timbre of a
musical instrument are immediately determined from a spectral analysis. The
color of a light source is determined by the spectrum of the electromagnetic
wave's electric field $E(t)$ as it fluctuates at an extremely high
frequency. Obtaining a spectrum from time series such as these involves the
Fourier transform, and generalizations based on Fourier analysis. 
The unit of power spectral density (PSD) is $I_{\mathrm{QD}}^{2}/\left(
R_{\mathrm{QD}}\omega \right) $. 

\begin{figure}
\includegraphics[width=85 mm]{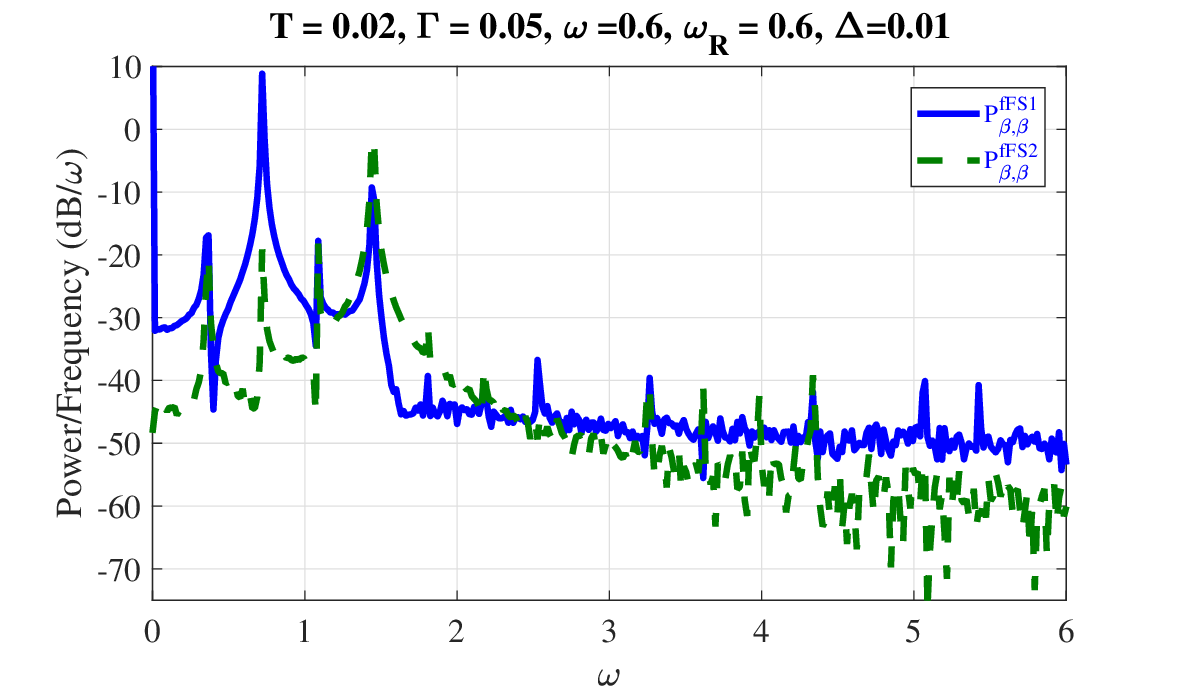} 
\caption{{The Power Spectral Density averaged over initial times near the primary resonance $%
\omega $ for $\varsigma =0.2$, detuning $\Delta =1$, and the EF frequences $\omega=0.1,$ $0.5$, $0.6$, and $0.8$. Here, $P^{\rm fFS1}_{\beta,\beta}$ and $P^{\rm fFS2}_{\beta,\beta}$ correspond to the active (i.e., $\Re e(I_{\rm RFAT})$) and reactive (i.e., $\Im m(I_{\rm RFAT})$) components of the electric current respectively.}}
\label{Fig_19}
\end{figure}

In Fig.~\ref{Fig_19} we present the computational results for the Power Spectral Density averaged over initial times near the primary resonance $\omega $ for $\varsigma =0.2$, detuning $\Delta =1$, and the EF frequences $\omega = 0.1,$ $0.5$, $0.6$, and $0.8$.
Similar plot in Fig.~\ref{Fig_24} shows the Power Spectral Density for for TLS in QD with Autler-Townes splitting. In Figs~\ref{Fig_19} and \ref{Fig_24}, the curves $P^{\rm fFS1}_{\beta,\beta}$ and $P^{\rm fFS2}_{\beta,\beta}$ correspond to the active (i.e., $\Re e(I_{\rm RFAT})$) and reactive (i.e., $\Im m(I_{\rm RFAT})$) components of the electric current respectively.

\begin{figure}
\includegraphics[width=85 mm]{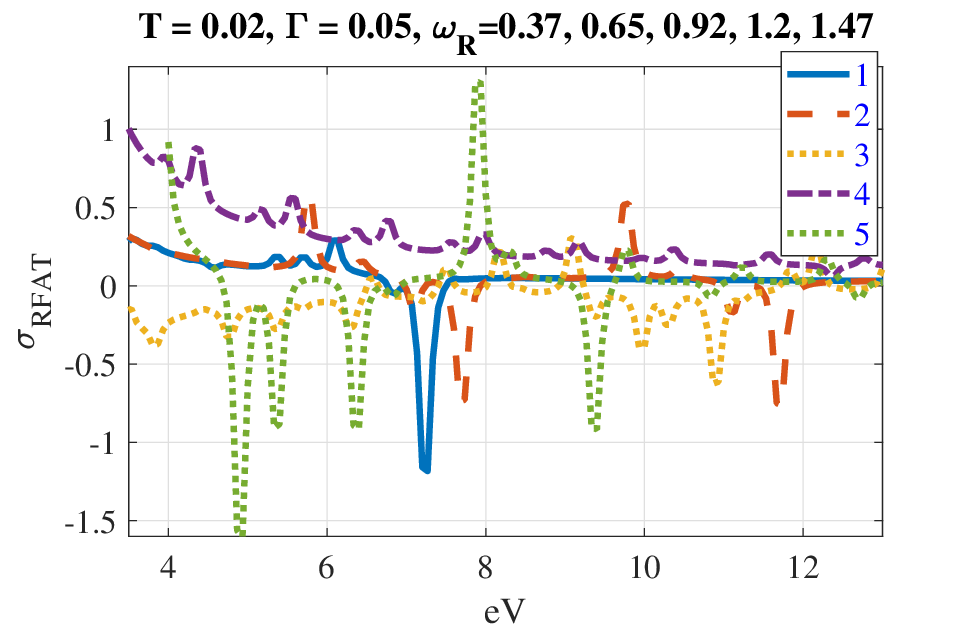} 
\caption{{The RFAT flop-assistent differential conductance $\sigma _{\mathrm{RFAT}}\left( eV_{%
\mathrm{bias}}\right) $ of QD for different Rabi frequencies $\omega_{R} = 0.37$, $0.65$, $0.92$, $1.2$, $1.47$.}}
\label{Fig_20}
\end{figure}

In Fig.~\ref{Fig_20} we show the RFAT flop-assistent differential conductance $\sigma _{\mathrm{RFAT}}\left( eV_{\mathrm{bias}}\right) $ of QD for different Rabi frequencies $\omega_{R} = 0.37$, $0.65$, $0.92$, $1.2$, and $1.47$.

\subsection{RFAT and Autler-Townes splitting}\label{sec3-7}

When EF is tuned in resonance (or close) to the transition frequency $\omega_Q $ of a given spectral line associated with quantized states in QD, one observes the Autler--Townes effect \cite{AutTow} (also known as the AC Stark effect or a dynamical Stark effect), causing changes in the shape of the absorption/emission spectra. 
Below, to take into account Autler--Townes effect \cite{AutTow}, we use a model involving QD with several quantized levels in the same QS. Additional energy levels originate from the ac Stark splitting resulting in the energy eigenvalues $E_{\pm }=-\hbar \Delta /2\pm \hbar \sqrt{\omega _{R}^{2}+\Delta ^{2}}/2$. The eigenstates of the QD system are dubbed $\left\vert +\right\rangle $ and $\left\vert -\right\rangle $.

The result of the ac field acting on QD is thus to shift the strongly coupled bare QD energy eigenstates into two states $\left\vert +\right\rangle $ and $\left\vert -\right\rangle $ and which are now separated by $\hbar \omega_{R} $. Evidence of this shift is obtained in the QD's I-V curve, which shows two shoulders around the steady-state position, separated by $\hbar \omega _{R}/e$ due to the Autler-Townes splitting. The conventional detuning $\Delta =\omega -\left( \varepsilon _{2}-\varepsilon _{1}\right) $ modifies as 
\begin{eqnarray}
\Delta _{1} &=&\Delta -\omega _{R}  \nonumber \\
\Delta _{2} &=&\omega _{R} \text{.}
\end{eqnarray}%

\begin{figure}
\includegraphics[width=85 mm]{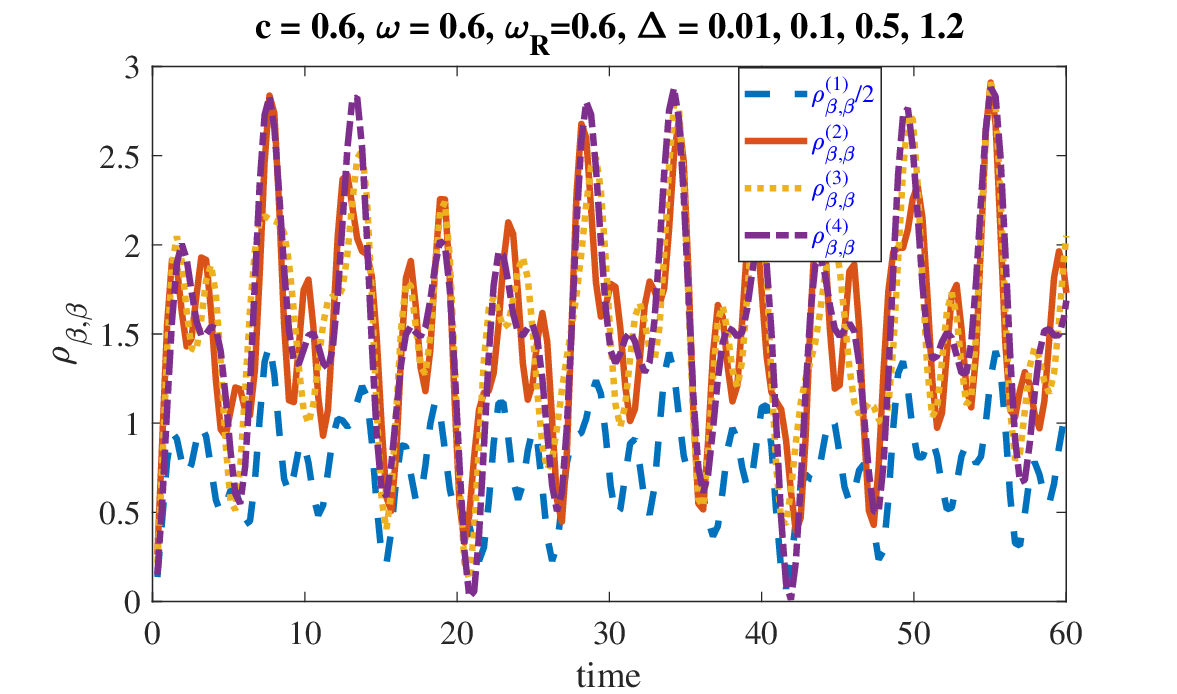} 
\caption{{Density matrix for TLS in QD with Autler-Townes splitting. Here $\varsigma = 0.6$, $\omega =0.6$, $\Delta =0.01$, $0.1$, $0.5 $, and $1.2$ for $\rho^{(1)}_{\beta,\beta }/2$, $\rho^{(2)}_{\beta,\beta }$, $\rho^{(3)}_{\beta,\beta }$, and $\rho^{(4)}_{\beta,\beta }$ respectively.}}
\label{Fig_21}
\end{figure}

The density matrix for TLS in QD with Autler-Townes splitting is presented in Fig~\ref{Fig_21} for $\varsigma = 0.6$, $\omega =0.6$, $\Delta =0.01$, $0.1$, $0.5 $, and $1.2$ for $\rho^{(1)}_{\beta,\beta }/2$, $\rho^{(2)}_{\beta,\beta }$, $\rho^{(3)}_{\beta,\beta }$, and $\rho^{(4)}_{\beta,\beta }$ respectively.

\begin{figure}
\includegraphics[width=85 mm]{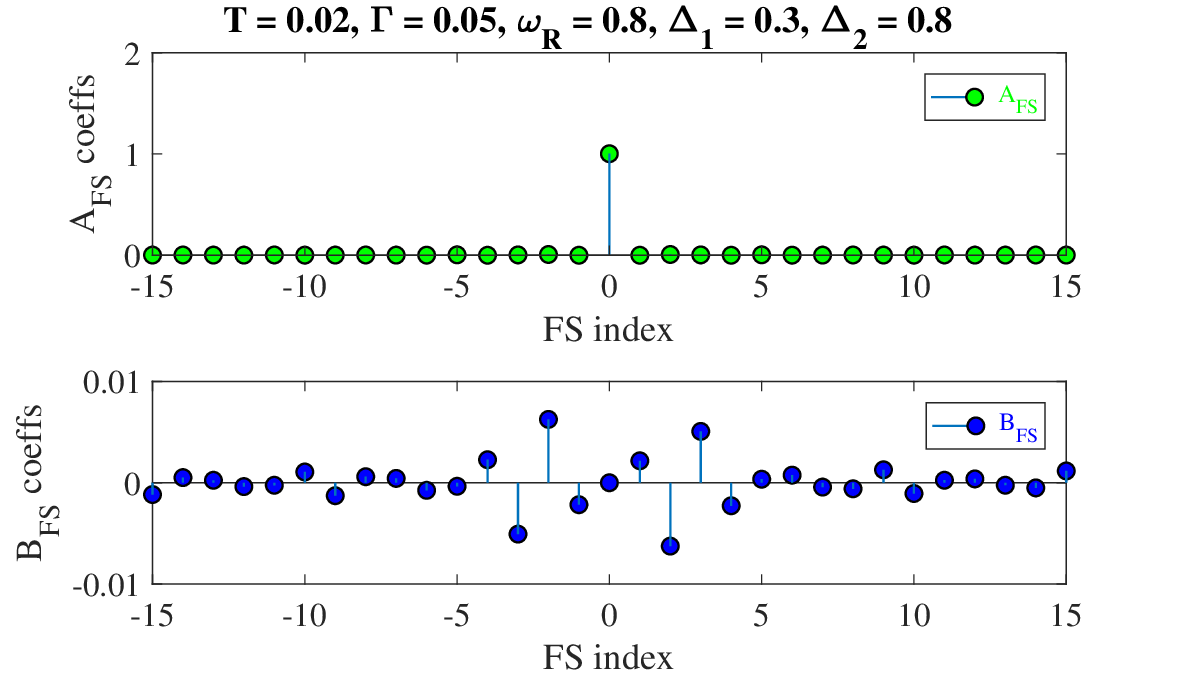} 
\caption{{The respective FS coefficients $A_n$ and $B_n$ related to the density matrix shown in the former Fig.~\ref{Fig_21}.}}
\label{Fig_22}
\end{figure}
\begin{figure}
\includegraphics[width=85 mm]{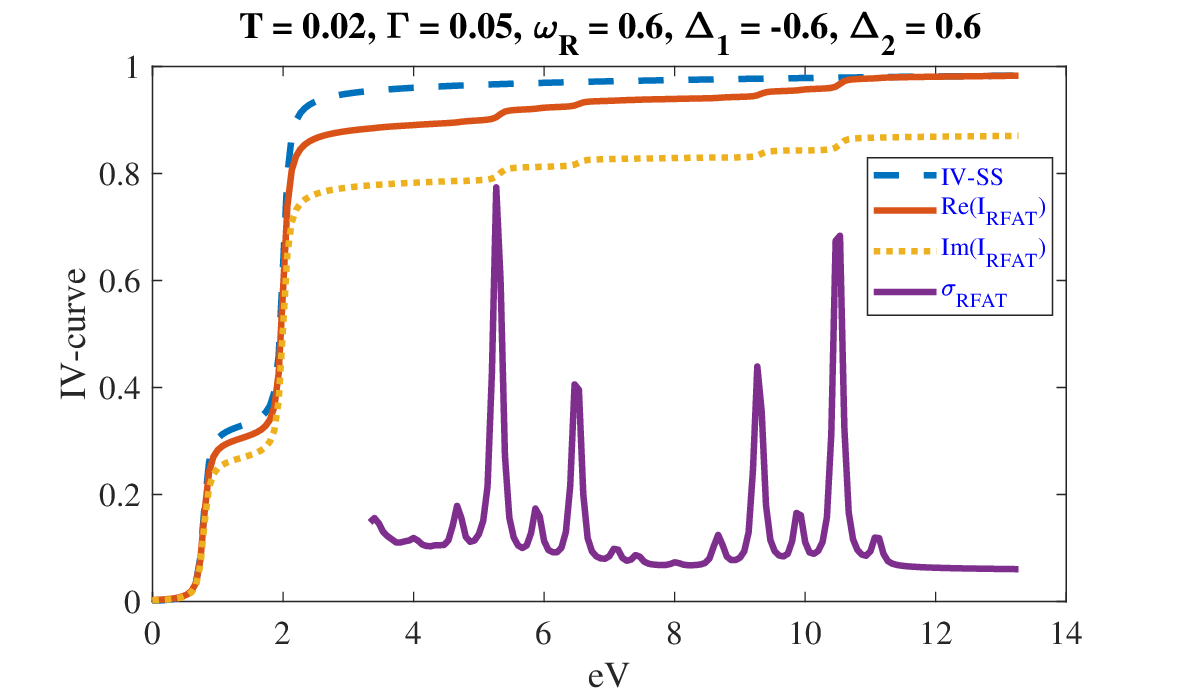} 
\caption{{The I-V curve for TLS in QD with Autler-Townes splitting.}}
\label{Fig_23}
\end{figure}
\begin{figure}
\includegraphics[width=85 mm]{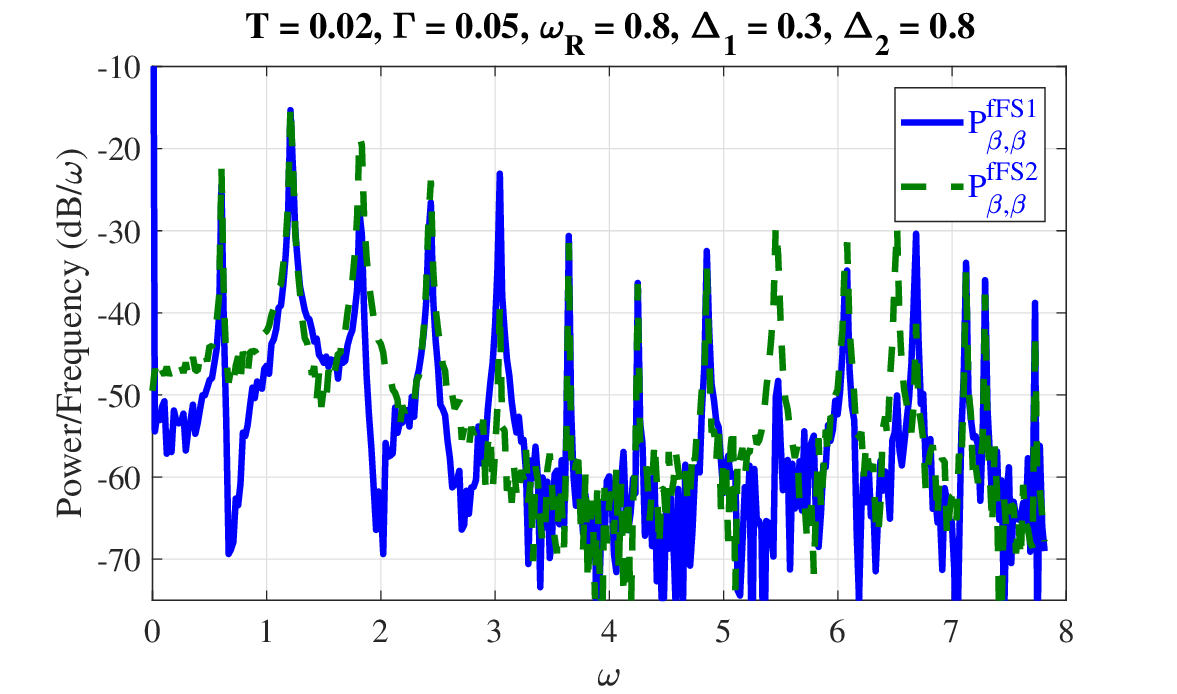} 
\caption{{The Power Spectral Density for for TLS in QD with Autler-Townes splitting. The curves $P^{\rm fFS1}_{\beta,\beta}$ and $P^{\rm fFS2}_{\beta,\beta}$ correspond to the active (i.e., $\Re e(I_{\rm RFAT})$) and reactive (i.e., $\Im m(I_{\rm RFAT})$) components of the electric current respectively.}}
\label{Fig_24}
\end{figure}

In Fig.~\ref{Fig_22} we present the Fourier series coefficients $A_n$ and $B_n$ corresponding to the density matrix shown in Fig.~\ref{Fig_21}. In Fig.~\ref{Fig_23} we show the respective I-V and $\sigma_{\rm RFAT}$ curves for TLS in QD with Autler-Townes splitting.

\section{Discussion}\label{sec5a}

The simplified model in the previous Sec.~\ref{sec3} allows a better understanding of the microscopic processes forming the electric current through QD with quantized states exposed to the external electromagnetic field. However, in an actual experiment, all the mechanisms are combined, resulting in a mixture of the conventional PAT, RFAT, and Coulomb blockade observed in the IV curves. Nevertheless, the performed analyses readily allow us to recognize and sort out the different mechanisms from each other. We exploit the fact, that the mentioned mechanisms cause the IV curves' ladders with different periods of the step spacing and heights. 

Let us discuss why the different $I_{\mathrm{QD}}\left( V\right) $ curves show various series of ascending and descending steps corresponding to peaks and dips in $\sigma _{\mathrm{RFAT}}\left( V\right) $. On the one hand, in some curves, the steps in $I_{\mathrm{QD}}\left( V\right) $, pronounced as peaks and dips in $\sigma _{\mathrm{RFAT}}\left( V\right) $, thus they are either are just ascending (see, e.g., Figs.~\ref{Fig_17}, \ref{Fig_23}) or just descending  (see, e.g., Figs.~\ref{Fig_12}, \ref{Fig_13}). On the other hand, in other $I_{\mathrm{QD}}\left( V\right) $ curves the steps change from ascending to descending back and forth  (see, e.g., Figs.~\ref{Fig_13}-- \ref{Fig_16}, \ref{Fig_18}). Our interpretation involves the multi-quantum transitions between the electron states. For instance, the three-quantum transition can be imagined as resulting from the alternate actions of the rotating and anti-rotating components of the oscillating field \cite{Hirschfelder1,Hirschfelder2,Shirley}. Using analogy with an electron in a magnetic field we can describe it as follows. When the electron initially is in the lower state $m=-1/2$,it first absorbs a photon from the rotating ($M_{z}=+1$) component of the field, going to the upper state $m=1/2$ with a non-conservation of energy. It cannot absorb another rotating photon because of angular momentum limitations, but it can absorb an antirotating ($M_{z}=-1$) photon to return virtually to state $m=-1/2$, but adding energy. Finally, it absorbs another rotating ($M_{z}=+1$) photon to arrive in the $m=1/2$ state with energy and angular momentum conservation. Since the energy-nonconserving intermediate states have very short lifetimes, a very high photon density is required to get all three photons in before the system decays back into the initial state. The classical solution of this problem will also show this resonance. In the experiments sketched in Figs.~\ref{Fig_1}, \ref{Fig_2}, measuring the electric current in QD, instead of a magnetic field one uses the gate and bias voltages.

One may also consider other multiple quantum transitions for the two-state system. The unperturbed energy associated with $\left\vert \alpha 0\right\rangle $ is $\left( -\omega +\omega _{Q}\right) /2$ and with $\left\vert \beta p\right\rangle $, $\left( 2p+1/2\right) \omega -\omega_{Q}/2$. These are approximately degenerate for $(2p+1)\omega =\omega _{Q}$. Thus multiple quantum transitions should occur for $\omega _{Q}$ equal to any odd harmonic of the driving frequency. They do not occur at even harmonics, because our perturbation has only off-diagonal elements concerning the atomic states. For the electron in a magnetic field, the selection rule can also be understood from the conservation of the $z$ component of angular momentum.

The $n$-quantum transition between two states, say $\alpha $ and $\beta $,
using that the resonant frequency is $\omega _{\mathrm{res}}=\left(
\varepsilon _{\beta }-\varepsilon _{\alpha }\right) /\hbar =\omega
_{Q}+\delta _{\alpha }-\delta _{\beta }$. One finds \cite{Salwen} the
transition probability as%
\begin{widetext}
\begin{equation}
P_{\beta \leftarrow \alpha }\left( t\right) =\rho _{\beta \beta }\left(
t\right) =\frac{\left\vert \varsigma _{n}\right\vert ^{2}}{\left( n\omega
-\omega _{\mathrm{res}}\right) ^{2}+\left\vert \varsigma _{n}\right\vert ^{2}%
}\sin ^{2}\left[ \sqrt{\left( n\omega -\omega _{\mathrm{res}}\right)
^{2}+\left\vert \varsigma _{n}\right\vert ^{2}}\frac{t}{2}\right]  \label{7}
\end{equation}%
\end{widetext}
where $\delta _{\alpha }$ and $\delta _{\beta }$ are level shifts due to
interactions with other states and $\frac{1}{2}\varsigma _{n}$ is the $n$-th
order matrix element connecting the two states. Eq. (\ref{7}) is the
generalization of the Rabi formula to multiple quantum transitions. The
above formulation reveals it to be valid for any transition where resonance
exists between only two levels. In discussing specific examples henceforth
we shall not write out Eq. (\ref{7}) but merely write down the resonance
frequency $\omega _{\mathrm{res}}$ and line width parameter $\varsigma _{n}$
to be substituted therein by comparing the 2$\times $2 matrix by
approximating as before.

Above we have studied the cases $p=0$ and $1$. In the same approximation we
also obtain the Rabi-type solutions for arbitrary $p$. The result is the
same as Eq. (\ref{7}) but with $n=2p+1$%
\begin{widetext}
\begin{equation}
P_{\beta \leftarrow \alpha }=\frac{\left\vert \varsigma _{p}\right\vert ^{2}%
}{\left[ \left( 2p+1\right) \omega -\omega _{\mathrm{res}}\right]
^{2}+\left\vert \varsigma _{p}\right\vert ^{2}}\sin ^{2}\left[ \sqrt{\left[
\left( 2p+1\right) \omega -\omega _{\mathrm{res}}\right] ^{2}+\left\vert
\varsigma _{p}\right\vert ^{2}}\frac{t}{2}\right]  \label{7b}
\end{equation}%
\end{widetext}
\begin{eqnarray}
\omega _{\mathrm{res}} &=&\omega _{Q}+\frac{2p+1}{p\left( p+1\right) }\frac{%
\varsigma ^{2}}{\omega }\text{ \ or \ }\omega _{Q}+\frac{\left( 2p+1\right)
^{2}}{p\left( p+1\right) }\frac{\varsigma ^{2}}{\omega _{Q}}\text{, \ \ }p>0
\nonumber \\
\varsigma _{p} &=&\frac{\varsigma ^{2p-1}}{2^{2p-1}\left( p!\right)
^{2}\omega ^{2p}}\text{ \ or \ \ }\frac{\left( 2p+1\right) ^{2p}}{%
2^{2p-1}\left( p!\right) ^{2}}\frac{\varsigma ^{2p+1}}{\left[ \omega _{Q}%
\right] ^{2p}}\text{ \ \ \ \ }p\geq 0  \label{37}
\end{eqnarray}%
where $\frac{1}{2}\varsigma $ again is the matrix element connecting the two
states. Keeping $\varsigma _{p}$ and $\omega $ fixed and looking at the
transition probability as a function of $\omega _{Q}$, Eq. (\ref{7b}) says
we should find resonances near $\omega _{Q}=\omega ,$ $3\omega ,$ $5\omega ,$
$7\omega ,...$ . The shifts are all towards smaller $\omega _{Q}$ and
decrease with increasing $p$. The widths decrease rapidly, because of the
powers of $c/\omega $. However even if $c/\omega $ were large, the widths
would eventually decrease because of the factorials in the denominator. So
only a finite number of resonances would have sufficient width to be
observable.

Otherwise, when $\varsigma $ and $\omega _{Q}$ are fixed we find the transition probability as a function of $\omega $. Eq. (\ref{7b}) predicts we should find resonances near $\omega =\omega _{Q},$ $\omega _{Q}/3,$ $\omega _{Q}/5,$ $\omega _{Q}/7,...$. The shifts are all towards a larger $\omega $. The widths decrease as $p$ increases only when $c/\omega _{Q}$ is small. The transition probability becomes simpler as a function of $\omega_{Q}$ than as a function of $\omega $. In the electron transport experiments, it is usually measured as a function of $\omega _{Q}$ by varying the
appropriate gate voltages.

When the interaction parameter $\varsigma $ is small and $\omega $ is almost equal to $\omega _{Q}$, the rotating-wave approximation (RWA) gives accurate predictions for the properties of the two-level system. The RWA consists of approximating $\mathcal{H}$ by the 2$\times $2 matrix 
\begin{equation}
\mathcal{H}_{\mathrm{RWA}}=\left( 
\begin{array}{cc}
-\omega _{Q}/2 & \varsigma \\ 
\varsigma & \omega _{Q}/2-\omega%
\end{array}%
\right) 
\begin{array}{c}
\leftarrow \alpha ,0 \\ 
\leftarrow \beta ,-1%
\end{array}%
\end{equation}%
The two eigenvalues of $\mathcal{H}_{\mathrm{RWA}}$ are the solutions of the
quadratic equation corresponding to the secular equation%
\begin{equation}
\left\vert \mathcal{H}_{\mathrm{RWA}}-W\mathbf{I}\right\vert =0
\end{equation}%
Thus%
\begin{equation}
W_{\pm }=\hbar \left[ -\omega /2\pm \varpi \right]
\end{equation}%
where%
\begin{equation}
\varpi =\sqrt{\Delta ^{2}+\varsigma ^{2}}\text{ \ \ \ \ \ }\Delta =\left(
\omega -\omega _{Q}\right) /2
\end{equation}%
For an $n$-quantum transition between two states, say $\alpha $ and $\beta $
we find 
\begin{equation}
P_{\beta \leftarrow \alpha }=\frac{\left\vert \upsilon \right\vert ^{2}}{%
\left[ n\omega -\omega _{\mathrm{res}}\right] ^{2}+\left\vert \upsilon
\right\vert ^{2}}\sin ^{2}\left[ \sqrt{\left[ n\omega -\omega _{\mathrm{res}}%
\right] ^{2}+\left\vert \upsilon \right\vert ^{2}}\frac{t}{2}\right]
\end{equation}%
but with%
\begin{eqnarray}
\omega _{\mathrm{res}} &=&\omega _{Q}+\frac{4n}{\left( n^{2}-1\right) }\frac{%
\varsigma _{n}^{2}}{\omega }  \nonumber \\
\upsilon &=&\frac{\varsigma _{n}^{n}}{2^{n-2}\left( \left( \left( n-1\right)
/2\right) !\right) ^{2}\omega ^{n-1}}
\end{eqnarray}

When computing the electric current, we exploited that in Floquet formalism, the time evolution can be completely absorbed into the rotating frame of reference. Frequently these can be done without employing Floquet's theorem. An advantage of the Floquet approach is that, even if there remains a time dependence in the rotating frame, one can still make the problem tractable using the periodicity of the Hamiltonian. For instance, it becomes beneficial when separating fast dynamics, within one period $T$ of the Hamiltonian, from slow ones, which change from one period to the next. Te suggested immediate experimental observation of the intrinsic features of such separation is useful for better understanding the EF-induced mixed state in quantum dots. 

\section{Conclusions}\label{sec4-c}

In conclusion, we studied how the Rabi flop-assisted tunneling is pronounced in the I-V curves and the differential conductance of quantum dots with a single TLS and when the Autler-Townes splitting occurs. We find that RFAT induces a series of steps and plateau in the I-V curves, whose position and spacing reflect the time dependence of the density matrix of QD as well as the intrinsic features of the EF-induced mixed quantum state. This opens the path to immediate experimental measurements of important intrinsic features of the mixed quantum state arising in the two-level system polarized by the external EF. Using RFAT promises novel opportunities for designing QD that exploits properties of the EF-induced mixed state to perform in a broad spectrum of EF. 

Summarizing, we find that ({\it i}) when studying PAT in the systems with quantized states, one should carefully check (a) the ladder spacing (i.e., the step length), which are different in the PAT and RFAT ladders as mentioned above and (b) the presence of descending steps, which depend on the ac field intensity and frequency.  ({\it ii}) The evaluations of the inter-level spacing and dipole coupling for the two states of particular interest for real samples are readily done by using the derived formulas. ({\it iii}) The suggested RFAT effect is quite general and can be pronounced in tiny junctions with quantized states. It is important to attach source and drain electrodes to the island with quantized levels whose spacing roughly matches the frequency of the external field with sufficient field intensity. ({\it iv})The relevant experiment could use junctions where the source and drain electrodes are attached to the island with quantized levels whose spacing roughly matches the frequency of the external field with intensity being sufficient to observe the RFAT ladder in the IV-curve. An example of such a device is shown in Figs.~\ref{Fig_1}, \ref{Fig_2}.

Although RFAT may be observed in arbitrary QD, promising design involves the narrow graphene stripes with zigzag atomic edges (see Figs.~\ref{Fig_1}, \ref{Fig_2}), where one may achieve an efficient intrinsic spectral narrowing of quantized levels by several orders of magnitude \cite{My-AQT,My-PRB23}. In such all-electrically tunable devices, one may considerably diminish the noise equivalent power (NEP) and improve the room-temperature performance in the broad spectral range.

Among various practical applications of RFAT, one example is finding parameters of the external electromagnetic field. In particular, by measuring the QD I-V curves, one extracts $\omega $ and $V_{ac}$ from the conventional PAT  features while from the complementary RFAT one also extracts $\omega _{\rm R}$. Additionally, by combining the PAT and RFAT data, one determines $\omega $, $\Delta $, ${\bf d}$, and $V_{\rm ac}$.

\section{Supplementary material}\label{sec4-d}

In Supplementary material (Sec.~\ref{Supplement}), we outline the Floquet formalism used to describe RFAT in QD, solve the time-dependent electron wave function equations, and compute the electron spectrum, density matrix, and PAT electric current.

In Sec.~B of Sec.~\ref{Supplement}, we present the expression for the density matrix of a two-level system. 

In Sec.~C of Sec.~\ref{Supplement}, we derive the PAT current through the quantum dot for arbitrary transparency of QD interfaces and non-stationary electron distribution inside QD. During the numeric computing of the RFAT ladder, we utilize analytical expressions for the density matrix given in Sec.~B. We also use the time-dependent electron distribution function $n_{i}\left( t\right)$ in the central quantum dot section C, which allows computing the respective electric current. These allow examining the effect of Rabi flops, which are expressed in the time dependence of $\delta n_{i}\left( t\right)$ and as ladders in the IV curves.

In Sec.~D of Sec.~\ref{Supplement}, we obtain the time-dependent electric current $I_{\rm QD}\left( t\right)$  (S30) and their expansion (S32) with the trigonometric coefficients $A_{0}$, $A_{n}$, and $B_{n}$ shown in plots Figs.~\ref{Fig_10}, \ref{Fig_22} of main text. We also compute the trigonometric Fourier series expansion of the electric current $I_{\rm QD}$ in QD. The obtained expressions (S36)--(S38) simplify the numeric calculations considerably.

\section{Supplementary material: Rabi flop-assisted electron tunneling through the quantum dots}\label{Supplement}

\subsection{Floquet Hamiltonian}\label{sec5-1}

The time-dependent Schr\"{o}dinger equation%
\begin{equation}
\mathcal{H}\left( \mathbf{r},t\right) \Psi \left( \mathbf{r},t\right) =0
\label{SE}
\end{equation}%
where%
\begin{equation}
\mathcal{H}\left( \mathbf{r},t\right) =H\left( \mathbf{r},t\right) -i\hbar 
\frac{\partial }{\partial t}
\end{equation}%
and $H\left( \mathbf{r},t\right) =H\left( \mathbf{r},t+\tau \right) $ is the
Hamiltonian which is periodic with the period $\tau =2\pi /\omega $ where $\omega $ is the angular frequency. The unperturbed Hamiltonian $H^{\left( 0\right) }\left( \mathbf{r}\right) $ has
a complete set of eigenstates $\alpha $, $\beta $, ... with orthonormal
eigenfunctions%
\begin{equation}
H^{\left( 0\right) }\left( \mathbf{r}\right) \alpha \left( \mathbf{r}\right)
=\varepsilon _{\alpha }^{\left( 0\right) }\alpha \left( \mathbf{r}\right) 
\text{, \ \ \ \ \ }\left\langle \beta \left( \mathbf{r}\right) \mid \alpha
\left( \mathbf{r}\right) \right\rangle =\delta _{\beta \alpha } \text{,} \label{SE0}
\end{equation}%
where $\varepsilon _{\alpha }^{\left( 0\right) }$ are the steady state eigenenergies.
The Hamiltonian $\mathcal{H}\left( \mathbf{r},t\right) $ is a Hermitian
operator in a combined space-time Hilbert space, $\mathbf{R}\times \mathbf{T}
$. The spatial part is spanned by the functions $\alpha \left( \mathbf{r}%
\right) $, $\beta \left( \mathbf{r}\right) $, .... The temporal part is
spanned by the complete orthogonal set of functions $e^{in\omega t}$ where $%
n=0,\pm 1,\pm 2$, .... Since $\mathcal{H}\left( \mathbf{r},t\right) $ is
Hermitian, it has a complete set of eigenstates $\chi _{k}\left( \mathbf{r},t\right)$%
\begin{equation}
\mathcal{H}\left( \mathbf{r},t\right) \chi _{k}\left( \mathbf{r},t\right)
=W_{k}\chi _{k}\left( \mathbf{r},t\right) \text{,}  \label{SW}
\end{equation}%
where $W_{k} $ are quoted as pseudoenergies.
Here the subscript $k$ corresponds to the $k$-th Floquet mode. The
eigenfunctions $\chi _{k}\left( \mathbf{r},t\right)$ of $\mathcal{H}\left( \mathbf{r},t\right) $ are periodic, $%
\chi _{k}\left( \mathbf{r},t\right) $ = $\chi _{k}\left( \mathbf{r},t+\tau
\right) $, and the pseudoenergies $W_{k}$ are real. It then follows that%
\begin{equation}
\Psi _{k}\left( \mathbf{r},t\right) =\chi _{k}\left( \mathbf{r},t\right)
e^{-iW_{k}t/\hbar }
\end{equation}%
is a solution of the Schrodinger equation (\ref{SE}).

For every Floquet mode k, there is an infinite number of related solutions $%
km$ of Eq. (\ref{SW}) which have%
\begin{equation}
W_{km}=W_{k}+m\hbar \omega \text{, \ \ \ \ }\chi _{km}\left( \mathbf{r}%
,t\right) =\chi _{k}\left( \mathbf{r},t\right) e^{im\omega t}
\end{equation}%
where $m=0,\pm 1,\pm 2$, .... However, not all of these solutions are
physically significant since $\Psi _{k}\left( \mathbf{r},t\right) =\Psi
_{km}\left( \mathbf{r},t\right) $ and the $km$ mode is physically equivalent
to the $k$-th mode. It follows that all of the physically different
solutions to the Schr\"{o}dinger equation have pseudoenergies $W_{k}$ lying within
an energy range of $\hbar \omega $. Thus, we limit our consideration to
Floquet modes $k$ for which%
\begin{equation}
-\hbar \omega <W_{k}<0
\end{equation}%
Within this range, there are the same number of Floquet modes as there are
independent solutions to the unperturbed Schr\"{o}dinger equation (\ref{SE0}%
). It follows that the $\chi _{k}\left( \mathbf{r},t\right) $ can be taken
to be orthogonal%
\begin{equation}
\left\langle \left\langle \chi _{k}\left( \mathbf{r},t\right) \mid \chi
_{k^{\prime }}\left( \mathbf{r},t\right) \right\rangle \right\rangle =0\text{
\ \ for }k^{\prime }\neq k
\end{equation}%
(where the double bracket means integration over all space and integration
over one period of time). Furthermore, for any initial conditions, $\Psi
\left( \mathbf{r},t\right) $ can be expressed as a linear combination of the 
$\Psi _{k}\left( \mathbf{r},t\right) $%
\begin{equation}
\Psi \left( \mathbf{r},t\right) =\sum_{k}C_{k}\Psi _{k}\left( \mathbf{r}%
,t\right)
\end{equation}%
Since both $\chi _{k}\left( \mathbf{r},t\right) $ and $H\left( \mathbf{r}%
,t\right) $ are periodic, they may be expanded in the basis sets $\alpha
\left( \mathbf{r}\right) $, $\beta \left( \mathbf{r}\right) $ ... and the $%
e^{in\omega t}$ functions%
\begin{eqnarray}
\left\langle \alpha \left( \mathbf{r}\right) \mid \chi _{k}\left( \mathbf{r}%
,t\right) \right\rangle &=&\sum_{n=-\infty }^{\infty }\left\langle \alpha
,n\mid \chi _{k}\right\rangle e^{in\omega t}  \label{nc1} \\
\left\langle \beta \left( \mathbf{r}\right) \left\vert H\left( \mathbf{r}%
,t\right) \right\vert \chi _{k}\left( \mathbf{r},t\right) \right\rangle
&=&\sum_{n=-\infty }^{\infty }H_{\beta \alpha }^{\left[ n\right]
}e^{in\omega t}  \label{nc2}
\end{eqnarray}%
Then, by multiplying eq (\ref{SW}) by $\beta \left( \mathbf{r}\right)
e^{-in\omega t}$, making use of the completeness equation%
\begin{equation}
\sum_{\alpha }\left\vert \alpha \left( \mathbf{r}\right) \right\rangle
\left\langle \alpha \left( \mathbf{r}\right) \right\vert =1
\end{equation}%
as well as eq (\ref{nc1}) and (\ref{nc2}), and integrating over all space
and one period of time, we obtain the \emph{Floquet equation}%
\begin{equation}
\sum_{\alpha ,n}\mathcal{H}_{\beta m,\alpha n}\left\langle \alpha ,n\mid
\chi \right\rangle =W\left\langle \beta ,m\mid \chi \right\rangle
\end{equation}%
where $\mathcal{H}$ is the \emph{Floquet Hamiltonian} which has the matrix
elements 
\begin{equation}
\mathcal{H}_{\beta m,\alpha n}=H_{\beta ,\alpha }^{\left[ m-n\right]
}+n\hbar \omega \delta _{\beta ,\alpha }\delta _{n,m}  \label{FH}
\end{equation}%
From eq (\ref{FH}) it follows that the $W_{k}$ are the solutions of the
secular equation%
\begin{equation}
\left\vert \mathcal{H-W}\mathbf{I}\right\vert =0
\end{equation}%
where $\mathbf{I}$ is the identity matrix. Furthermore, the $\left\langle
\alpha ,n\mid \chi _{k}\right\rangle $ are the components of the eigenvector
of $\mathcal{H}$ corresponding to $W_{k}$.

\subsection{Density matrix for the two-level system}\label{sec5-2}

The general expression for the density matrix can be specialized
to the case of a two-level system. Taking the initial condition to be $\rho
_{\alpha \alpha }(t_{0}\equiv 0)=1$, we obtain%
\begin{widetext}
\begin{eqnarray}
\rho _{\beta \beta }\left( t\right) &=&2\sum_{k}\left\langle \alpha ,2k\mid
\chi _{+}\right\rangle \sum_{l}\left\langle \alpha ,2l\mid \chi
_{-}\right\rangle \times \cos 2\omega_{\rm R}t\sum_{mn}\cos \left[ 2\left( n-m\right) \omega t\right]
\left\langle \beta ,2m+1\mid \chi _{+}\right\rangle \left\langle \beta
,2n+1\mid \chi _{-}\right\rangle  \nonumber \\
&&-\sin 2\omega_{\rm R}t\sum_{mn}\sin \left[ 2\left( n-m\right) \omega t\right]
\left\langle \beta ,2m+1\mid \chi _{+}\right\rangle \left\langle \beta
,2n+1\mid \chi _{-}\right\rangle  \nonumber \\
&&+\left\vert \sum_{k}\left\langle \alpha ,2k\mid \chi _{+}\right\rangle
\right\vert ^{2}\sum_{mn}\cos \left[ 2\left( n-m\right) \omega t\right]
\left\langle \beta ,2m+1\mid \chi _{+}\right\rangle \left\langle \beta
,2n+1\mid \chi _{+}\right\rangle  \nonumber \\
&&+\left\vert \sum_{k}\left\langle \alpha ,2k\mid \chi _{-}\right\rangle
\right\vert ^{2}\sum_{mn}\cos \left[ 2\left( n-m\right) \omega t\right]
\left\langle \beta ,2m+1\mid \chi _{-}\right\rangle \left\langle \beta
,2n+1\mid \chi _{-}\right\rangle \text{,}  \label{A1}
\end{eqnarray}%
where $\omega_{\rm R} $ is the Rabi frequency. For the two-level system (TLS) $\omega_{\rm R} = \sqrt{\Delta^2 + \varsigma^2}$, where $\Delta = \omega - \omega_{\rm R}$ is the detuning and $\varsigma $ is the inter-level coupling strength.
\begin{eqnarray}
\rm{Re}\rho _{\alpha \beta }\left( t\right) &=&\sum_{k}\left\langle \alpha
,2k\mid \chi _{+}\right\rangle \sum_{l}\left\langle \alpha ,2l\mid \chi
_{-}\right\rangle  \nonumber \\
&&\times \cos 2\omega_{\rm R}t\sum_{mn}\cos \left[ \left( 2n-2m-1\right) \omega t\right]
\times \left[ \left\langle \alpha ,2n\mid \chi _{-}\right\rangle
\left\langle \beta ,2m+1\mid \chi _{+}\right\rangle +\left\langle \alpha
,2n\mid \chi _{+}\right\rangle \left\langle \beta ,2m+1\mid \chi
_{-}\right\rangle \right]  \nonumber \\
&&-\sin 2\omega_{\rm R}t\sum_{mn}\sin \left[ \left( 2n-2m-1\right) \omega t\right] \times %
\left[ \left\langle \alpha ,2n\mid \chi _{-}\right\rangle \left\langle \beta
,2m+1\mid \chi _{+}\right\rangle -\left\langle \alpha ,2n\mid \chi
_{+}\right\rangle \left\langle \beta ,2m+1\mid \chi _{-}\right\rangle \right]
\nonumber \\
&&+\left\vert \sum_{k}\left\langle \alpha ,2k\mid \chi _{+}\right\rangle
\right\vert ^{2}\sum_{mn}\cos \left[ \left( 2n-2m-1\right) \omega t\right]
\left\langle \alpha ,2n\mid \chi _{+}\right\rangle \left\langle \beta
,2m+1\mid \chi _{+}\right\rangle  \nonumber \\
&&+\left\vert \sum_{k}\left\langle \alpha ,2k\mid \chi _{-}\right\rangle
\right\vert ^{2}\sum_{mn}\cos \left[ \left( 2n-2m-1\right) \omega t\right]
\left\langle \alpha ,2n\mid \chi _{-}\right\rangle \left\langle \beta
,2m+1\mid \chi _{-}\right\rangle  \label{A2}
\end{eqnarray}
\end{widetext}
The long-time averages of the density matrix elements are defined as%
\begin{equation}
\left\langle \rho _{\gamma \gamma ^{\prime }}\right\rangle
=\lim_{T\rightarrow \infty }\frac{1}{T}\int_{0}^{T}\rho _{\gamma \gamma
^{\prime }}\left( t\right) dt  \label{A3}
\end{equation}%
Performing the averaging over initial times near the primary resonance $\omega $ (see Ref.~\cite{Hirschfelder1}) gives%
\begin{eqnarray}
\left\langle \rho _{\beta \beta }\right\rangle &=&\left\vert
\sum_{k}\left\langle \alpha ,2k\mid \chi _{+}\right\rangle \right\vert
^{2}\sum_{l}\left\vert \left\langle \beta ,2l+1\mid \chi _{+}\right\rangle
\right\vert ^{2}  \nonumber \\
&&+\left\vert \sum_{k}\left\langle \alpha ,2k\mid \chi _{-}\right\rangle
\right\vert ^{2}\sum_{l}\left\vert \left\langle \beta ,2l+1\mid \chi
_{-}\right\rangle \right\vert ^{2}  \label{A4}
\end{eqnarray}%
We also have $\left\langle \rho _{\alpha \beta }\right\rangle =0$, $\rho
_{\alpha \alpha }=1-\rho _{\beta \beta }$, $\rm{Im}\left( \rho _{\alpha
\beta }\right) $ is not required, and we refer to $\left\langle \rho _{\beta
\beta }\right\rangle $ as the "saturation".

\subsection{Electric current through QD}\label{sec5-0}

The knowledge of density matrix (see Sec.~\ref{sec5-2} above) and of the time-dependent electron distribution function $n_{i}\left( t\right)$ in the central quantum dot section C (see Figs.~1, 2 and Sec.~IIIB in main text) allows computing the respective electric current in QD. We would like to examine the effect of Rabi flops, which are expressed in the time dependence of $\delta n_{i}\left( t\right)$. Instead of using the tunneling Hamiltonian model, whose applicability is limited to the limit of low-transparent interface barrier, we use the Octavio-style boundary conditions \cite{Octavio1,Landauer,Buttiker,Octavio2}, which are valid for arbitrary transparency of QD interfaces \cite{Datta-1995} and take into account the non-stationary electron distribution inside QD. The electric tunneling current $I_{\mathrm{QD}}$ in QD is obtained \cite{Landauer,Buttiker} as 
\begin{equation}
I_{\mathrm{QD}}=\frac{2e}{h}\int_{-\infty }^{\infty }d\varepsilon \left[
n_{\rightarrow }\left( \varepsilon \right) -n_{\leftarrow }\left(
\varepsilon \right) \right] \text{.} \label{I-dc}
\end{equation}%
Let us consider QD in the piece-wise geometry of the double-barrier junction NICIN, where N denotes the electrode, I is the barrier, and C is the central section of QD. We assume that the bias voltage $V$ drops on the two equal barriers I, so the voltage drop on each of the barriers is $V/2$. If $v_{\mathrm{F}}\tau _{\varepsilon }>>d_{\mathrm{C}}$, where $v_{\mathrm{F}}$ is the Fermi velocity, $\tau _{\varepsilon }$ is the
energy relaxation time, $d_{\mathrm{C}}$ is the central section length, the
distribution functions $n_{\rightarrow }\left( \varepsilon \right) $ and $%
n_{\leftarrow }\left( \varepsilon \right) $ entering Eq.~(\ref{I-dc}) in the
central section deviate from the equilibrium Fermi function. Defenitely, this is the case for graphene QD with $d_{\rm C} = 5$~nm, where $v_{\mathrm{F}} \sim 10^6$~m/s, and $\tau _{\varepsilon } \sim 10^{-12} - 10^{-6}$~s. When the finite bias voltage $V$ is applied to the NICIN junction, within the central section C one gets 
\begin{equation}
n_{\rightleftharpoons }\left( E,L\right) \simeq n_{\rightleftharpoons
}\left( E,0\right) \text{; but }n_{\rightarrow }\left( E,0\right) \neq
n_{\leftarrow }\left( E,0\right) \text{.}  \label{elast}
\end{equation}%
The functions $n_{\rightarrow }\left( E,x\right) $ and $n_{\leftarrow
}\left( E,x\right) $ in the central section C are related to the distribution
functions in the electrodes by the Octavio-type boundary conditions \cite%
{Datta-1995,Octavio1,Octavio2}.

When, additionally, the ac field of frequency $\omega $ is applied, one obtains the boundary conditions (BC) for the electron distribution function in the central section C of the NICIN junction in the form
\begin{equation}
n_{\rightarrow }^{\rm C}\left( \varepsilon ,0\right) =\sum_{n}R_{n}\left(
\varepsilon \right) n_{\leftarrow }^{\rm C}\left( \varepsilon ,0\right)
+\sum_{n}T_{n}\left( \varepsilon \right) n_{0}\left( \varepsilon -n\omega
+eV/2\right)  \label{BC1}
\end{equation}%
and%
\begin{equation}
n_{\leftarrow }^{\rm C}\left( \varepsilon ,L\right) =\sum_{m}R_{m}\left(
\varepsilon \right) n_{\rightarrow }^{\rm C}\left( \varepsilon ,L\right)
+\sum_{m}T_{m}\left( \varepsilon \right) n_{0}\left( \varepsilon -m\omega
-eV/2\right)   \label{BC2} \text{.}
\end{equation}%
The partial reflection $R_{n}$ and transmission $T_{n}$ coefficients entering Eqs.~(\ref{BC1}), (\ref{BC2}) are obtained by matching the electron wavefunctions at the interfaces as described in Refs.~\cite{My-AQT,My-PRB23}. Thus, we get two types of BC: one type is the Octavio-type boundary conditions \cite{Datta-1995,Octavio1,Octavio2} (\ref{BC1}), (\ref{BC2}) for the electron distribution functions $n_{\rightleftharpoons } $ while the other set of BC is needed for matching the electron wavefunctions at the interfaces in piece-wise geometry as in Refs.~\cite{My-AQT,My-PRB23}. First, we use the method \cite{My-AQT,My-PRB23} to solve the boundary conditions for the electron wavefunctions for QD in the piece-wise geometry. Since the electron properties in section C of QD are described by Floquet formalism we compute how the electron eigenenergies and eigenvectors depend on the EF amplitude $E_{\mathrm{ac}}$ and frequency $\omega $. Now we use Floque Hamiltonian in place of the steady-state Hamiltonian  used in the former works \cite{My-AQT,My-PRB23}. The obtaned electron wavefunctions allow microscopic computing of the dipole matrix elements $\mathbf{d}=\left\langle i\right\vert e\mathbf{r}\cdot \mathbf{E}_{\mathrm{ac}}\left\vert j\right\rangle $ characterizing the EF-induced mixing of two quantum states $\left\vert i\right\rangle $ and $\left\vert j\right\rangle $ due to their coupling with each other in QD positioned at the radius-vector $\mathbf{r}$. In this way, we calculate the EF-dependent electron reflection $R_{n}$ and transmission $T_{n}$ probabilities, which are then substituted to the above expressions for $n_{\rightleftharpoons }^{\rm C}$ in Eqs.~(\ref{BC1}), (\ref{BC2}).
In Eqs.~(\ref{BC1}), (\ref{BC2}), we use that the change of the electron distribution
function $\delta n\left( \varepsilon, t\right) $ in C due to the non-stationary Rabi flops is%
\[
n_{\rightleftharpoons }^{\rm C}\left( \varepsilon ,L\right)
-n_{\rightleftharpoons }^{\rm C}\left( \varepsilon ,0\right)  = \delta n_{\rm C} (t)  \text{.%
} 
\]%
We also take into account that $\delta n_{\rm C} (t) $ is positive for the electrons
propagating from the left to the right end of the central section as well as
for those propagating in the opposite direction. Then%
\begin{eqnarray}
n_{\leftarrow }^{\rm C}\left( \varepsilon ,0\right) &=&n_{\leftarrow }^{\rm C}\left(
\varepsilon ,L\right) -\delta n_{\rm C}  \nonumber \\
n_{\rightarrow }^{\rm C}\left( L\right) &=&n_{\rightarrow }^{\rm C}\left( 0\right)
-\delta n_{\rm C}
\end{eqnarray}%
\begin{widetext}
\begin{eqnarray*}
n_{\rightarrow }^{\rm C}\left( 0\right) &=&\sum_{n}R_{n}\left( \varepsilon
\right) n_{\leftarrow }^{\rm C}\left( \varepsilon ,0\right) +\sum_{n}T_{n}\left(
\varepsilon \right) n_{0}\left( \varepsilon -n\omega +eV/2\right) \\
&=&R_{1}^{2}n_{\rightarrow }^{\rm C}\left( L\right)
+R_{1}\sum_{n}T_{n}n_{0}\left( \varepsilon -n\omega -eV/2\right)
+\sum_{n}T_{n}\left( \varepsilon \right) n_{0}\left( \varepsilon -n\omega
+eV/2\right) +R_{1}\delta n_{\rm C} \\
&=&R_{1}^{2}n_{\rightarrow }^{\rm C}\left( 0\right)
+R_{1}\sum_{n}T_{n}n_{0}\left( \varepsilon -n\omega -eV/2\right)
+\sum_{n}T_{n}\left( \varepsilon \right) n_{0}\left( \varepsilon -n\omega
+eV/2\right) +\left( R_{1}-R_{1}^{2}\right) \delta n_{\rm C}\text{,}
\end{eqnarray*}%
or
\[
n_{\rightarrow }^{\rm C}\left( 0\right) =\frac{1}{1-R_{1}^{2}}\left(
\sum_{n}T_{n}\left[ R_{1}n_{0}\left( \varepsilon -n\omega -eV/2\right)
+n_{0}\left( \varepsilon -n\omega +eV/2\right) \right] +R_{1}\left(
1-R_{1}\right) \delta n_{\rm C}\right) \text{,} 
\]%
\end{widetext}
where we introduced the notation
\begin{equation}
R_{1}\left( \varepsilon \right) =\sum_{n}R_{n}\left( \varepsilon \right) 
\text{.}
\end{equation}%
In the limit $V=0$, we get%
\[
n_{\rightarrow }^{\rm C}\left( 0\right) =\frac{1}{1-R_{1}}\left(
\sum_{n}T_{n}n_{0}\left( \varepsilon -n\omega \right) +R_{1}\delta
n^{\rm C}\right) \text{.} 
\]%
In the above formulas we use
\begin{eqnarray}
n_{\leftarrow }^{\rm C}\left( L\right) &=&n_{\leftarrow }^{\rm C}\left( 0\right)
+\delta n_{\rm C}  \nonumber \\
n_{\rightarrow }^{\rm C}\left( L\right) &=&n_{\rightarrow }^{\rm C}\left( 0\right)
-\delta n_{\rm C}
\end{eqnarray}%
\begin{widetext}
\begin{eqnarray*}
n_{\leftarrow }^{\rm C}\left( \varepsilon ,L\right) &=&\sum_{m}R_{m}\left[
n_{\rightarrow }^{\rm C}\left( 0\right) -\delta n_{\rm C}\right] +\sum_{m}T_{m}n_{0}%
\left( \varepsilon -m\omega -eV/2\right) \\
&=&\frac{1}{1-R_{1}^{2}}\sum_{n}T_{n}\left[ n_{0}\left( \varepsilon -n\omega
-eV/2\right) +R_{1}n_{0}\left( \varepsilon -n\omega +eV/2\right) \right] +\frac{R_{1}^{2}-R_{1}-1}{R_{1}+1}\delta n_{\rm C} \ \text{.}
\end{eqnarray*}%
Finally we get%
\[
n_{\rightarrow }^{\rm C}\left( 0\right) =\frac{1}{1-R_{1}^{2}}\left(
\sum_{n}T_{n}\left[ R_{1}n_{0}\left( \varepsilon -n\omega -eV/2\right)
+n_{0}\left( \varepsilon -n\omega +eV/2\right) \right] +R_{1}\left(
1-R_{1}\right) \delta n_{\rm C}\right)  \text{,}
\]%
\[
n_{\leftarrow }^{\rm C}\left( 0\right) =\frac{1}{1-R_{1}^{2}}\sum_{n}T_{n}\left[
n_{0}\left( \varepsilon -n\omega -eV/2\right) +R_{1}n_{0}\left( \varepsilon
-n\omega +eV/2\right) \right] +\frac{R_{1}^{2}-2R_{1}-2}{R_{1}+1}\delta
n^{\rm C} (t)   \text{,}
\]%
and
\begin{equation}
n_{\rightarrow }^{\rm C}\left( 0\right) -n_{\leftarrow }^{\rm C}\left( 0\right) =%
\frac{1}{1+R_{1}}\left( \sum_{n}T_{n}\left[ n_{0}\left( \varepsilon -n\omega
+eV/2\right) -n_{0}\left( \varepsilon -n\omega -eV/2\right) \right]
+R_{1}\delta n_{\rm C}\right)  \text{,}
\end{equation}%
\end{widetext}
where we used%
\[
n_{\rightarrow }^{\rm C}\left( \varepsilon ,L\right) -n_{\rightarrow }^{\rm C}\left(
\varepsilon ,0\right) =\delta n_{\rm C}   \text{,}
\]%
and%
\begin{eqnarray}
n_{0}\left( \varepsilon \right) &=&\frac{1}{\exp \left( \varepsilon
/T\right) +1}   \text{.}
\end{eqnarray}%
In the limit $V = 0$, and if we neglect the contribution from Rabi flops, we
simply get%
\[
n_{\leftarrow }^{\rm C}\left( 0\right) =\frac{R_{1}\left( R_{1}-2\right) }{%
R_{1}-1}\sum_{n}T_{n}n_{0}\left( \varepsilon -n\omega \right)   \text{,}
\]   %
\[
n_{\leftarrow }^{\rm C}\left( \varepsilon ,L\right) =n_{\rightarrow }^{\rm C}\left(
\varepsilon ,0\right) =\frac{1}{1-R_{1}\left( \varepsilon \right) }%
\sum_{n}T_{n}\left( \varepsilon \right) n_{0}\left( \varepsilon -n\omega
\right)      \text{.}
\]%
Under the last assumption, when the bias voltage $V = 0$, the averaged electric current vanishes 
\begin{equation}
I_{\rm QD} (V=0) = \frac{2e}{h}\int_{-\infty }^{\infty }d\varepsilon \left[ n_{\rightarrow
}^{\rm C}\left( \varepsilon \right) -n_{\leftarrow }^{\rm C}\left( \varepsilon
\right) \right] =0   \label{I-QD0} \text{,}
\end{equation}
where we used that $T_{n}=1-R_{n}$.  Then, the averaged electric current $I_{\mathrm{QD}}$ through QD becomes%
\begin{widetext}
\begin{eqnarray}
I_{\rm QD} (V) &=&\frac{2e}{h}\frac{1}{1+R_{1}}\int_{-\infty }^{\infty }d\varepsilon \left(
\sum_{n}T_{n}\left[ n_{0}\left( \varepsilon -n\omega +eV/2\right)
-n_{0}\left( \varepsilon -n\omega -eV/2\right) \right] +R_{1}\delta
n^{\rm C} (t)\right)  \label{IV-QD} 
 \text{,}
\end{eqnarray}%
\end{widetext} where $n^{\rm C} (t) $ is the time-dependent part of the electron distribution function. The last expression is consistent with the expression for the photon-assisted electric current in the two-barrier NICIN junction \cite{Dayem-Martin,Tien-Gordon} where the transmission probability is%
\[
T\left( \varepsilon \right) \rightarrow \sum_{nm}J_{n}^{2}\left( \alpha
_{1}\right) J_{m}^{2}\left( \alpha _{2}\right) T_{\mathrm{2B}}\left(
\varepsilon -n\omega -m\omega \right)    \text{.}
\]%
Introducing the partial barrier transpariencies $T_{1}$ and $T_{2}$ in the dc case one gets 
\[
T_{\mathrm{2B}}=T_{1}T_{2}\frac{1}{\left( 1-\sqrt{\left( 1-T_{1}\right)
\left( 1-T_{2}\right) }\right) ^{2}}e^{2i\delta _{\mathrm{C}}}   \text{,}
\]%
where the electron wavefunction phaseshifts on the C section and barriers is 
\begin{equation}
\delta _{\mathrm{C}}=\left( k_{\rm F}\pm \frac{V}{\hbar v_{F}}\pm \frac{%
\pi }{4}\frac{C_{\rm C}V_{g}}{e}\right) d_{\rm C}   \text{,}  \label{phase}
\end{equation}%
where in graphene $k_{\rm F}=0.77\pi /a\simeq 10^{10}$ m$^{-1}$, $v_{F} \sim 10^{6}$\thinspace m/s, $C_{\rm C}$ is the C section capacitance,  $d_{\rm C}$ is the QD section length,  $V$ is the net voltage drop across the whole NICIN junction, and $V_{g}$ is the bottom gate voltage. 
For the Schottky barriers, the probabilities of reflection $R_n$ and of transmission $T_n$ along with a finite phase shift $\delta $ (which appears due to a finite barrier thickness) are extracted from the normal state experimental curves. In Eq. (\ref{phase}), the charging effect is taken into 
account by the third term in the round brackets. One can see that the QD capacitance actually determines the scale on which the gate voltage affects the whole junction conductance.

The dc currents with and without microwaves are related as%
\begin{equation}
I_{\mathrm{QD}}=\sum_{m=-\infty }^{\infty }a_{m}I_{\mathrm{QD}}^{\left(
0\right) }\left( eV_{\mathrm{dc}}+m\hbar \omega \right) \label{I-PAT}   \text{,}
\end{equation}%
where $I_{\mathrm{QD}}^{\left( 0\right) }$ is the current without radiation. The expansion coefficients $a_{m} $ of the time-dependent part of the distribution function $\delta n_{i}\left( t\right) $ into the Fourier series are computed in the next Sec.~\ref{sec5-0B}.
We see that the rectified current of QD biased by the voltage $V(t)=V_{\mathrm{dc }} + V_{\mathrm{ac}}\cos \omega t$ is given as the sum of the appropriately weighted dc-currents $I_{\mathrm{QD}}^{\left(0\right)} $ that are evaluated without ac driving and at voltages shifted by integer multiples of photon energies. 

\subsection{Rabi flop-assisted tunneling through quantum dot}\label{sec5-0B}

Let us consider QD exposed to EF, whose effect is depicted using the electric circuit in Fig.~2 of main text. In Fig.~2 of main text we show the schematics where the gate voltages $V_{\rm lg}$ are applied to the local gate electrodes of QD to control the heights of potential barriers separating the sections of QD formed on the graphene stripe. The external EF, by acting on the local gates, induces a potential difference $V_{\mathrm{ac} }\cos \Omega t$ between the central and adjacent sections of QD.  Neglecting the interaction of the microwave field with the barrier and considering one of the adjacent sections as a reference (e.g., the left region), the effect of the microwave field is to add a potential $V_{\mathrm{ac}}\cos \Omega t$ to the dc voltage $V_{\rm bias}$ in the central section. 
Importantly, within this simple model, the effect of the external field is accounted for by adding a time-dependent, but spatially homogeneous potential within the central section which is described by a local Hamiltonian%
\begin{equation}
\mathcal{H}=\mathcal{H}_{0}+eV_{\mathrm{ac}}\cos \Omega t
\end{equation}%
It is obvious that the time-dependent homogeneous potential does not vary the spatial distribution of the electronic wave function within each region. Solving the time-dependent Schr\"{o}dinger equation using the Floquet formalism, the time-dependent electron wavefunction for the central section can be written as%
\begin{equation}
\left\vert \alpha ,t\right\rangle =\left\vert \alpha \right\rangle
\sum_{m=-\infty }^{\infty }C_{m}e^{-im\Omega t}  \label{ksiRF}
\end{equation}%
where the sum in Eq. (\ref{ksiRF}) accounts for the Rabi flops between the two quantized states $\alpha $ and $\beta $.

Using the results of the preceding Sections we are now in a position to elucidate the effect of Rabi flops on the I-V curve of QD. The simplest way is to use the "reverse engineering". First, we compute the time-dependent density matrix, which is then used to find $\delta n_{i}\left( t\right)  =\mathrm{Tr}\left[ \delta \rho \left( t\right)  N_{i}\right] $. We expand the time-dependent part of the distribution function $\delta n_{i}\left( t\right) $ into the Fourier series using Eq.~(\ref{ksiRF}) and replacing $\Omega \rightarrow \tilde{\omega}_{\rm R}$, where $\tilde{\omega}_{\rm R} = \sqrt{\omega^2_{\rm R} + \Delta^2}$ is the generalized Rabi frequency. Then, the time-averaged distribution function, $\overline{n_{i}}^t  $, for the central region C can be written in terms of the steady-state distribution function, $n_{i}^{\left( 0\right) } $, without external potential as%
\begin{equation}
\overline{n_{i}}^t  = n_{i}^{\left( 0\right) }+\overline{\delta n_{i}\left( t\right)}^t = \sum_{m=-\infty
}^{\infty }a_{m}n_{i}^{\left( 0\right) }\left( \varepsilon +m\hbar \tilde{\omega}_{\rm R} \right) \text{.}
\label{Av}
\end{equation}%
The above Eq. (\ref{Av}) can be interpreted physically as follows: photon absorption ($%
m>0$) and emission ($m<0$) can be viewed as creating the effective electron pseudoenergy levels $\varepsilon \pm m\hbar \tilde{\omega}_{\rm R} $ with a probability given by $a_{m}$. 

The time-dependent electric current is expanded as
\begin{equation}
I_{\rm QD}\left( t\right) =\sum_{n=-\infty }^{\infty }C_{n}e^{in\tilde{\omega}_{\rm R}t}=A_{0}+\sum_{n=1}^{\infty }A_{n}\cos \left( n \tilde{\omega}_{\rm R} t\right) +B_{n}\sin \left( n\tilde{\omega}_{\rm R} t\right)  \text{,}
\end{equation}
where%
\begin{eqnarray}
A_{0} &=&C_{0}=\frac{1}{T}\int_{0}^{T}I_{\rm QD}\left( t\right) dt \text{,}  \nonumber \\
A_{n} &=&C_{n}+C_{-n}=\frac{2}{T}\int_{0}^{T}I_{\rm QD}\left( t\right) \cos \left( n\tilde{\omega}_{\rm R} t\right) dt  \text{,} \nonumber \\
B_{n} &=&i\left( C_{n}-C_{-n}\right) =\frac{2}{T}\int_{0}^{T}I_{\rm QD}\left(
t\right) \sin \left( n\tilde{\omega}_{\rm R} t\right) dt \text{.}
\label{IV-t}
\end{eqnarray}%
By evaluating the trigonometric coefficients $A_{0}$, $A_{n}$, and $B_{n}$ (see plots in Fig.~10 of main text) we compute the trigonometric Fourier series expansion of the electric current $I_{\rm QD}$ in QD. 
Now, using Eqs.~(\ref{I-dc}), (\ref{IV-QD}), we compute the electric current in QD where the RFAT is pronounced.
To obtain the active component of the time-averaged electric current, we substitute the obtained $\overline{n_{i}}^t $ into the expressions for the electric current derived in Section~\ref{sec5-0}. Finally, similar to Eq.~(\ref{Av}), we get
\begin{equation}
\overline{I_{\rm QD}(t)}^t  =  \sum_{n=-\infty
}^{\infty }A_{n}I_{\rm QD}^{\left( 0\right) }\left( \varepsilon +n\hbar \tilde{\omega}_{\rm R} \right) \text{,}
\label{IV-Av}
\end{equation}%
From Eq. (\ref{I-PAT}), it can be observed that tunneling between the
central and adjacent sections can happen from states of energy $\varepsilon $
in the left region to states of energy $\varepsilon \pm m\hbar \tilde{\omega}_{\rm R} $ in
the central section, namely through inelastic tunneling.

\end{document}